\documentclass[11pt,a4paper]{article}

\usepackage[margin=1in]{geometry}

\usepackage[utf8]{inputenc} 
\usepackage[T1]{fontenc}    
\usepackage{hyperref}       
\usepackage{url}            
\usepackage{booktabs}       
\usepackage{amsfonts}       
\usepackage{nicefrac}       
\usepackage{microtype}      
\usepackage{amsmath}
\usepackage{mathtools}
\usepackage{threeparttable}
\usepackage{graphicx}
\usepackage{listings}
\usepackage{xcolor}

\definecolor{codegreen}{rgb}{0,0.6,0}
\definecolor{codegray}{rgb}{0.5,0.5,0.5}
\definecolor{codepurple}{rgb}{0.58,0,0.82}
\definecolor{backcolour}{rgb}{0.95,0.95,0.92}
\lstdefinestyle{mystyle}{
    backgroundcolor=\color{backcolour},   
    commentstyle=\color{codegreen},
    keywordstyle=\color{magenta},
    numberstyle=\tiny\color{codegray},
    stringstyle=\color{codepurple},
    basicstyle=\ttfamily\footnotesize,
    breakatwhitespace=false,         
    breaklines=true,                 
    captionpos=b,                    
    keepspaces=true,                 
    numbers=left,                    
    numbersep=5pt,                  
    showspaces=false,                
    showstringspaces=false,
    showtabs=false,                  
    tabsize=4
}
\usepackage{multirow} 

\usepackage{authblk}
\providecommand{\keywords}[1]{\vspace{0.5cm}\noindent\textbf{\textit{Keywords:}} #1}

\title{\textbf{GeoFWI3D: Large-scale 3D Velocity Model Dataset for Deep Learning-assisted Seismic Imaging}}

\author[1]{Sujith Swaminadhan}
\author[1]{Yiran Shen}
\author[1]{Chao Li}
\author[2]{Kai Gao}
\author[2]{Ting Chen}
\author[1]{Sergey Fomel}
\author[1]{Tolulope Agbaje}
\author[3]{Yang Cui}
\author[3]{Liuqing Yang}
\author[4]{Jaewook Lee}
\author[1]{Robin Dommisse}
\author[5]{Umair bin Waheed}
\author[1]{Mrinal K. Sen}
\author[1]{Yangkang Chen}

\affil[1]{The University of Texas at Austin}
\affil[2]{Los Alamos National Laboratory}
\affil[3]{Uppsala University}
\affil[4]{The University of Texas at Permian Basin}
\affil[5]{King Fahd University of Petroleum and Minerals}

\begin{document}
\maketitle
\begin{abstract} \normalsize

We introduce GeoFWI3D, a large-scale open-source benchmark dataset of geologically plausible 3D subsurface models designed to accelerate deep learning (DL) assisted seismic imaging and full waveform inversion (FWI). FWI is a physics-driven, wave-equation-based optimization technique used in seismic imaging to estimate the subsurface properties. However, traditional FWI is highly non-linear, non-unique, and ill-posed. The iterative process often converges to local minima when the starting model is insufficiently accurate. Most importantly, 3D FWI is prohibitively expensive. With the advent of DL, a direct mapping between the shot gathers and subsurface properties can be established by training a network on realistic models. The primary bottleneck for such approaches is the lack of large-scale, realistic training datasets.  GeoFWI3D addresses this gap with 10,000 geologically diverse velocity models at 96$\times$96$\times$96 resolution, spanning four structural complexity classes: pure stratigraphy, faulted networks, salt diapirism, and complex coupled fault-salt systems. Each model is accompanied by co-registered multi-modal labels including compressional velocity (Vp), zero-offset seismic reflectivity, relative geologic time (RGT), and semantic fault/salt masks. To facilitate systematic evaluation, we introduce benchmark tasks covering fault detection, joint salt body segmentation and chronostratigraphy prediction, FWI, wavefield and traveltime surrogates using neural operators, and generative modeling with a 3D diffusion model. We present baseline results for each task to establish reference performance metrics for future users. The dataset is publicly available under Creative Commons Attribution 4.0 International.

\end{abstract}


\vfill

\section{Background \& summary}
Full waveform inversion (FWI) has emerged as one of the most powerful high-resolution seismic imaging methods for reconstructing subsurface physical properties from recorded seismic wavefields \cite{tarantola1984inversion}. By iteratively minimizing the misfit between observed and simulated seismic data using wave-equation-constrained optimization, FWI can recover detailed velocity structures critical for hydrocarbon exploration, geothermal characterization, carbon sequestration monitoring, earthquake studies, and subsurface engineering applications. Despite its remarkable imaging capability, conventional physics-driven FWI remains computationally demanding, highly nonlinear, and sensitive to acquisition incompleteness, noise contamination, and inaccurate initial models \cite{fichtner2010full}.

Recent advances in deep learning (DL) have introduced new opportunities to accelerate and stabilize seismic imaging and inversion workflows. DL-assisted methods have been explored for seismic inversion for direct velocity inversion \cite{araya2018deep,shucai2020}, seismic super-resolution \cite{li2021super}, seismic interpretation \cite{wu2019faultseg3d}, and surrogate wavefield modeling \cite{moseley2020solving}. More recently, diffusion models, neural implicit representations, and physics-informed learning frameworks have demonstrated significant potential to improve inversion robustness and to incorporate geological priors into seismic imaging workflows \cite{rasht2022physics,dhara2022physics,dhara2023elastic,taufik2024learned}. These developments are driving a transition from purely physics-driven inversion toward hybrid physics-AI seismic imaging paradigms. The success of pure data-driven modern DL methods strongly depends on the availability of large-scale, diverse, and geologically realistic training datasets. In computer vision, benchmark datasets such as ImageNet \cite{deng2009imagenet} have fundamentally transformed the development of deep neural networks by enabling standardized training and fair comparison among algorithms \cite{deng2022openfwi}. In contrast, seismic inversion and imaging research still lack sufficiently large and standardized benchmark datasets, especially for realistic three-dimensional problems. Early DL-based FWI studies often relied on simplified synthetic datasets constructed from horizontally layered models or small-scale variants of the Marmousi and SEG/EAGE salt models \cite{araya2018deep}. Although useful for proof-of-concept demonstrations, these datasets do not adequately represent the structural complexity and geological diversity encountered in real subsurface environments.

To address this limitation, several benchmark datasets have recently been proposed for seismic machine learning (ML) research. OpenFWI introduced a large-scale 2D synthetic seismic benchmark containing multiple geological scenarios and acquisition settings for data-driven FWI research \cite{deng2022openfwi}. OpenFWI significantly improved the reproducibility and comparability of DL-based inversion methods and became one of the first widely adopted benchmark datasets for seismic AI research. However, the geological realism of OpenFWI remains relatively limited because many of its models were generated primarily from procedural or structurally simplified synthetic scenarios. Motivated by the need for more geologically meaningful seismic benchmarks, the GeoFWI project was subsequently proposed to provide structurally more realistic velocity models for DL-assisted seismic imaging and inversion \cite{geofwi}. Compared with OpenFWI, GeoFWI emphasizes geological plausibility and structural diversity by incorporating folded stratigraphy, fault systems, salt-related deformation, unconformities, and more realistic sedimentary architectures. Rather than focusing solely on large-scale synthetic generation, GeoFWI aims to bridge the gap between simplified benchmark datasets and realistic geological structures encountered in practical seismic exploration and Earth imaging applications. As a result, GeoFWI provides a more challenging and geologically representative benchmark for evaluating the generalization capability, robustness, and interpretability of modern DL-based inversion algorithms.

Nevertheless, existing benchmark datasets, including both OpenFWI and GeoFWI, remain predominantly two-dimensional, whereas real seismic wave propagation and practical seismic imaging problems are fundamentally three-dimensional. The transition from two-dimensional to three-dimensional seismic imaging is not merely an increase in dimensionality, but rather a fundamental change in physical realism, computational complexity, and geological representation. In realistic Earth media, seismic waves propagate in fully three-dimensional space, producing out-of-plane scattering, multipathing effects, azimuthal illumination variations, and complex wavefield interactions that cannot be adequately represented in 2D approximations. Moreover, many important geological structures, including channel systems, fracture corridors, reef complexes, fault networks, and salt diapirs, are intrinsically three-dimensional. Consequently, DL models trained exclusively on 2D datasets often exhibit limited generalization capability when applied to realistic 3D seismic imaging problems. At the same time, modern seismic AI research is rapidly evolving toward three-dimensional architectures, including 3D convolutional neural networks, transformers, diffusion probabilistic models, Fourier neural operators, and neural implicit representations. These methods require large and diverse volumetric datasets for effective training, benchmarking, and reproducibility. The increasing availability of GPU-accelerated wave-equation solvers and high-performance computing platforms has further enabled the generation and utilization of large-scale 3D seismic datasets. Therefore, developing standardized and geologically realistic 3D benchmark datasets has become increasingly important for advancing next-generation AI-assisted seismic imaging technologies.

In this work, we introduce GeoFWI3D, a large-scale three-dimensional velocity model dataset designed for DL-assisted seismic imaging, inversion, and interpretation research. GeoFWI3D extends the GeoFWI philosophy from 2D to 3D by emphasizing geological realism, structural diversity, reproducibility, and computational scalability. The dataset contains 10,000 geologically reasonable 3D velocity models with volumetric dimensions of 96$\times$96$\times$96, including diverse structural patterns such as folded layers, faults, salt bodies, stratigraphic deformation, and complex subsurface heterogeneity. Compared with existing 2D datasets, GeoFWI3D provides substantially richer three-dimensional geological variability and more realistic seismic wave propagation environments for modern AI-based seismic imaging research.

In addition to the velocity models themselves, GeoFWI3D provides associated examples and workflows for multiple downstream geophysical applications. First, we provide examples that demonstrate the use of GeoFWI3D for both conventional physics-based and data-driven FWI, enabling direct comparison between physics-driven and AI-assisted inversion paradigms. Second, we demonstrate the use of GeoFWI3D for seismic interpretation tasks, including training DL models to predict geological features such as salt delineation, fault detection, and relative geological time (RGT) prediction. Third, we provide a flexible benchmark framework for evaluating modern and state-of-the-art DL methods in terms of imaging efficacy, inversion accuracy, generalization capability, and computational scalability. By integrating seismic imaging, inversion, and interpretation into a unified benchmark ecosystem, GeoFWI3D aims to facilitate reproducible research and accelerate the development of robust, generalizable, and physics-consistent AI methods for future subsurface imaging.

\section{Methods}

\begin{figure}
\centering
\includegraphics[width=0.9\columnwidth]{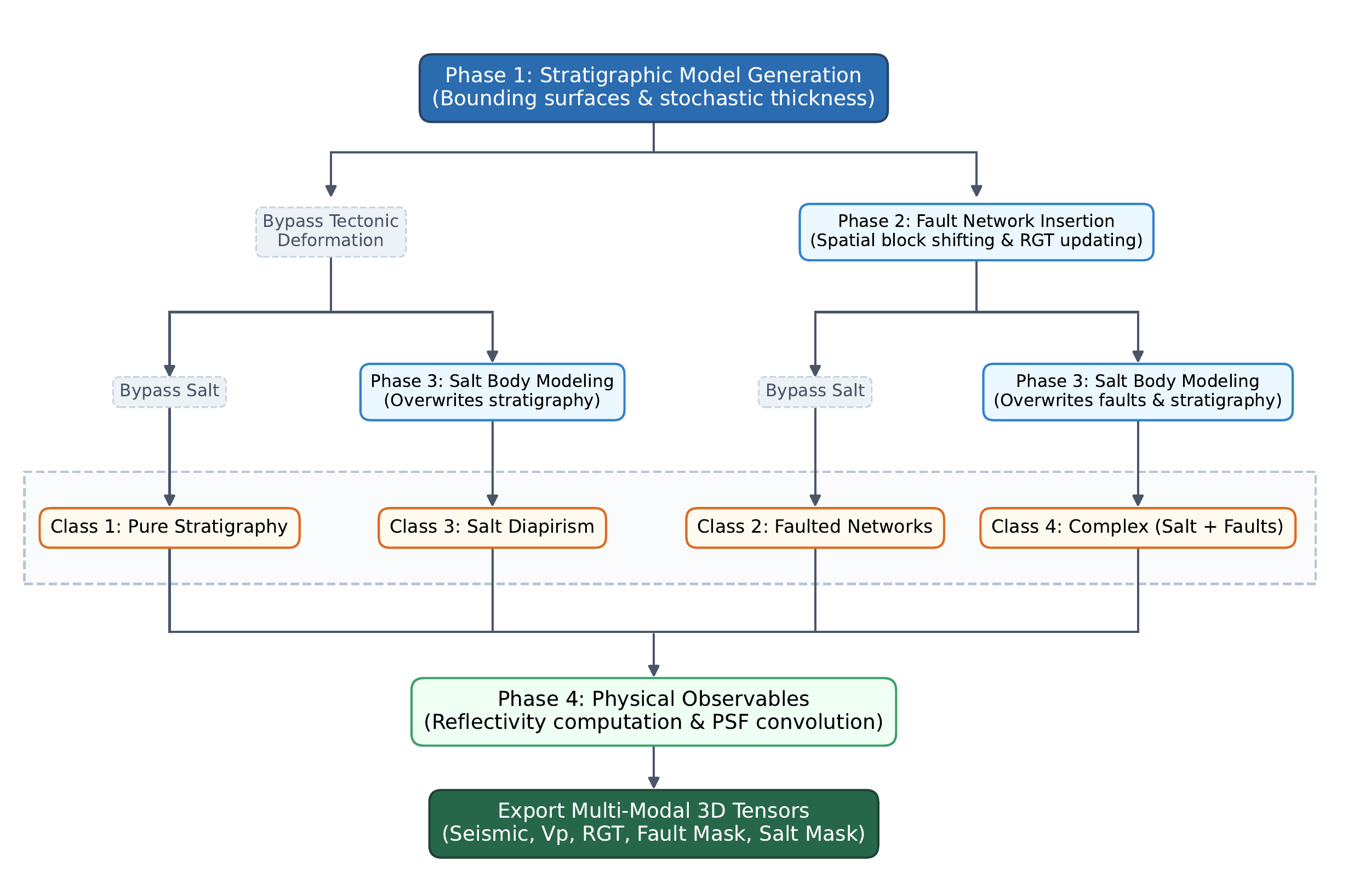}
\caption{The sequential flowchart for generating the distinct dataset classes. The workflow operates in four primary phases, beginning with the baseline generation of continuous stratigraphy (Phase 1). By stochastically bypassing the mechanical deformation of fault insertion (Phase 2) and the geometric modeling of salt diapirs (Phase 3), the strategy yields four distinct dataset classes of increasing geological complexity: pure stratigraphy, faulted networks, salt diapirism, and coupled salt and fault systems. All branches ultimately funnel into translating the structural kinematics into physical observables (Phase 4), finally exporting aligned multi-modal 3D tensors for each volume.}
\label{fig:workflow}
\end{figure} 

\begin{figure}
    \centering
    \includegraphics[width=0.98\linewidth]{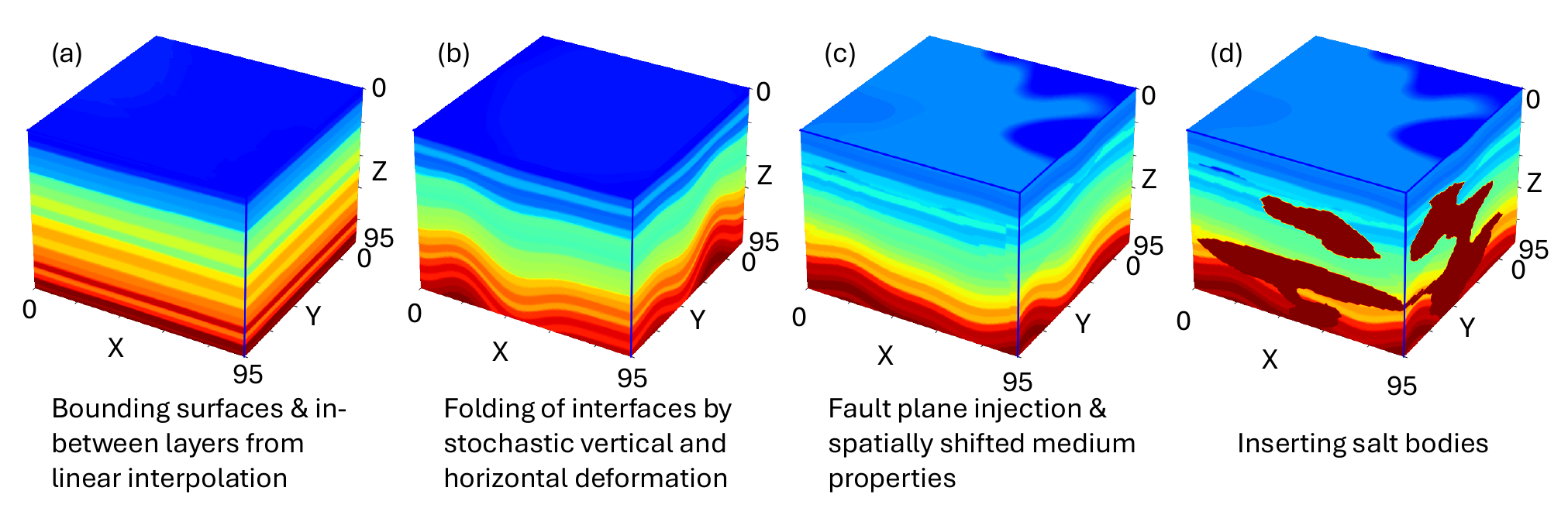}
    \caption{3D schematic representation of the strategy for generating subsurface models, operating in four sequential phases: (a) generating bounding surfaces and interfaces via linear interpolation (b) simulating sedimentary folding by applying stochastic vertical and horizontal deformations to the interfaces (c) injecting fault planes and applying spatial shifts to the medium properties and (d) inserting complex 3D salt bodies that overwrite the pre-existing stratigraphy.}
    \label{fig:workflow-schematic}
\end{figure}

To overcome scalability bottlenecks in ML for seismic imaging, we employed a procedural modeling strategy based on the open-source random geological model (RGM) framework \cite{GAO2026106133}. This multi-randomization approach enables efficient generation of diverse, high-fidelity, realistic 3D synthetic geological models featuring unconformities, complex fault networks, and salt bodies. The generation workflow operates in four primary phases, as demonstrated in Figure \ref{fig:workflow}, and the 3D schematic representation is shown in Figure \ref{fig:workflow-schematic}.
While the following subsections detail the specific parameterization and application of this pipeline to generate our dataset, the comprehensive mathematical derivation of the core RGM algorithm is detailed by Gao and Chen \cite{GAO2026106133}.

\subsection{Stratigraphic model generation}
Within the RGM framework \cite{GAO2026106133}, the initial unfaulted stratigraphic models are constructed by defining top and bottom bounding surfaces using randomized functions, such as random white noise surfaces, random Gaussian surfaces, Cauchy surfaces, or multi-octave Perlin \cite{etherington2022perlin} noise. To ensure geological realism, 2D Gaussian smoothing filters are applied to these sampled functions. The intermediate interfaces are calculated by linear interpolation between the bounding surfaces and subsequently adjusted using stochastic vertical and horizontal thickness variations. This approach mimics geological phenomena such as differential deposition and compaction, resulting in geologically informed, non-uniform, wavy sedimentary layering.
To populate the model with P-wave velocities ($V_p$), each sedimentary layer is assigned a distinct, constant random velocity value. Because layer interfaces rarely align perfectly with discrete integer grid points, assigning neighboring values can create "staircase" artifacts along the boundaries. To mitigate this, regularly sampled sub-grid points are generated strictly within each layer's boundaries and assigned the layer's designated velocity. These non-uniformly distributed spatial velocity values are then linearly interpolated along the depth axis onto the final regular computational mesh. A piecewise smooth RGT volume is also generated simultaneously by assuming a linearly increasing geological time from newer (top) to older (bottom) reflectors.

\subsection{Fault Network Insertion}
To simulate structurally realistic tectonic deformations, discrete faults are injected into the continuous field. The RGM framework employs a spatial-shifting approach in which grid points within a designated fault block are displaced. To encompass diverse structural styles, fault injection is stochastically partitioned into two tectonic regimes: an extensional regime featuring a single fault family with random strikes, and a conjugate regime featuring intersecting fault families with strikes grouped approximately $80^\circ$ to $100^\circ$ apart. 
Depending on the assigned complexity class, between 4 and 12 discrete faults are generated in each model. The framework supports both planar and listric geometries. Listric faults are defined by depth-dependent dip angles with a maximum deviation of up to $40^\circ$. Additional randomized parameters include strike, rake, dip, and spatial displacements ranging from 3 to 20 grid units. When a fault is inserted, the spatial shift concurrently modifies the $V_p$ model, the fault index model, and the RGT volume to ensure consistent structural alignment across all multi-modal datasets.

\subsection{Salt Body Modeling}
Complex 3D geobodies, such as upward-intruding salt diapirs, are introduced into the models using a four-step geometric procedure. For volumes designated to contain salt, 1 to 5 distinct salt bodies are generated. First, salt boundaries are initialized as randomized closed circular curves at specific control depths, with random phases in the Fourier domain. To ensure natural asymmetry, explicit path and radius variations are applied, with base radii scaled between 6 and 21 grid points to fit the computational domain. 
Second, these control curves are connected along the depth axis using a piecewise cubic Hermite interpolating polynomial (PCHIP) interpolation \cite{fritsch1980monotone}, which preserves the monotonicity and smoothness of the 3D salt profile. Third, a multi-octave Perlin noise surface is generated to serve as a random top-cutting surface, terminating the salt bodies at a relative depth between 40 and 70\% of the total vertical dimension. Finally, grid points enclosed within this defined 3D boundary are filled with a constant salt velocity ($v_{salt}$) sampled from 4480~m/s to 4530~m/s, thereby structurally overwriting the pre-existing stratigraphy and faults.

\subsection{Bandlimited Seismic Volume Generation}
To translate the structural kinematics into physical observables, synthetic reflectivity volumes are generated. First, continuous acoustic impedance models are derived from the generated $V_p$ volumes (density estimated using Gardner's rule \cite{gardner1974formation}), from which 3D analytical reflection coefficients are computed at each voxel interface, producing a raw, broadband reflectivity grid.  
The broadband reflectivity volume is convolved with a 3D spatial point spread function (PSF) \cite{claerbout1985imaging,bleistein2001mathematics} to generate bandlimited seismic image-like volumes. The PSF serves as a spatial-blurring kernel representing the imaging system's impulse response to a point scatterer.

To bypass the computational overhead of running full-wavefield finite-difference modeling and reverse time migration (RTM) across 10,000 individual volumes \cite{lecomte2008resolution,lecomte2015ray}, this convolution is executed efficiently in the 3D wavenumber domain ($\mathbf{k}$-space). Applying the convolution theorem, the final spatial image $I(\mathbf{x})$ is generated by multiplying the 3D Fast Fourier transform ($\mathcal{F}$) of the reflectivity $R(\mathbf{x})$ with the transformed PSF, followed by an inverse transform:
$$I(\mathbf{x}) = \mathcal{F}^{-1} \{ \mathcal{F}\{R(\mathbf{x})\} \cdot \mathcal{F}\{\text{PSF}(\mathbf{x})\} \}.$$
While these zero-offset acoustic reflectivity volumes
provide immediate utility for structural interpretation and segmentation tasks, FWI inherently requires raw, pre-stack wavefield recordings. Because optimal acquisition parameters vary significantly with the specific inversion objective, we provide the raw velocity volumes as the foundational physical medium rather than imposing a fixed acquisition footprint. This flexibility allows users to independently simulate custom 3D shot gathers tailored to their specific geometric requirements, such as varying source-receiver densities, maximum offsets, and frequency bandwidths. By deploying standard finite-difference (FD) or finite-element (FE) wave-equation solvers over the provided structural grids, researchers can generate the exact pre-stack data necessary to benchmark their specialized FWI algorithms.

\subsection{Parameter Space and Stochastic Sampling}
To ensure the generated dataset contains a comprehensive and statistically representative variety of geological scenarios, the RGM input parameters were constrained within realistic bounds and sampled using the Latin hypercube sampling (LHS) \cite{mckay2000comparison,Stein01051987}. LHS guarantees near-uniform coverage across the high-dimensional parameter space, preventing the clustering and gaps common in standard Monte Carlo randomization \cite{morris1995exploratory}.
The spatial dimensions of the generated models were standardized to a $96 \times 96 \times 96$ computational grid. Hence, the spatial geometric parameters were linearly scaled from the RGM settings to match this volume. To mimic distinct tectonic regimes, fault generation was split into two topological styles. An extensional regime characterized by a single fault family with randomly distributed strikes, and a conjugate regime characterized by two intersecting fault families with strikes grouped approximately $90^\circ$ apart. Unconformities were stochastically introduced into approximately 67\% of the generated models, varying between shallow and deep erosional boundaries.

\begin{figure}
\centering
\includegraphics[width=0.99\columnwidth]{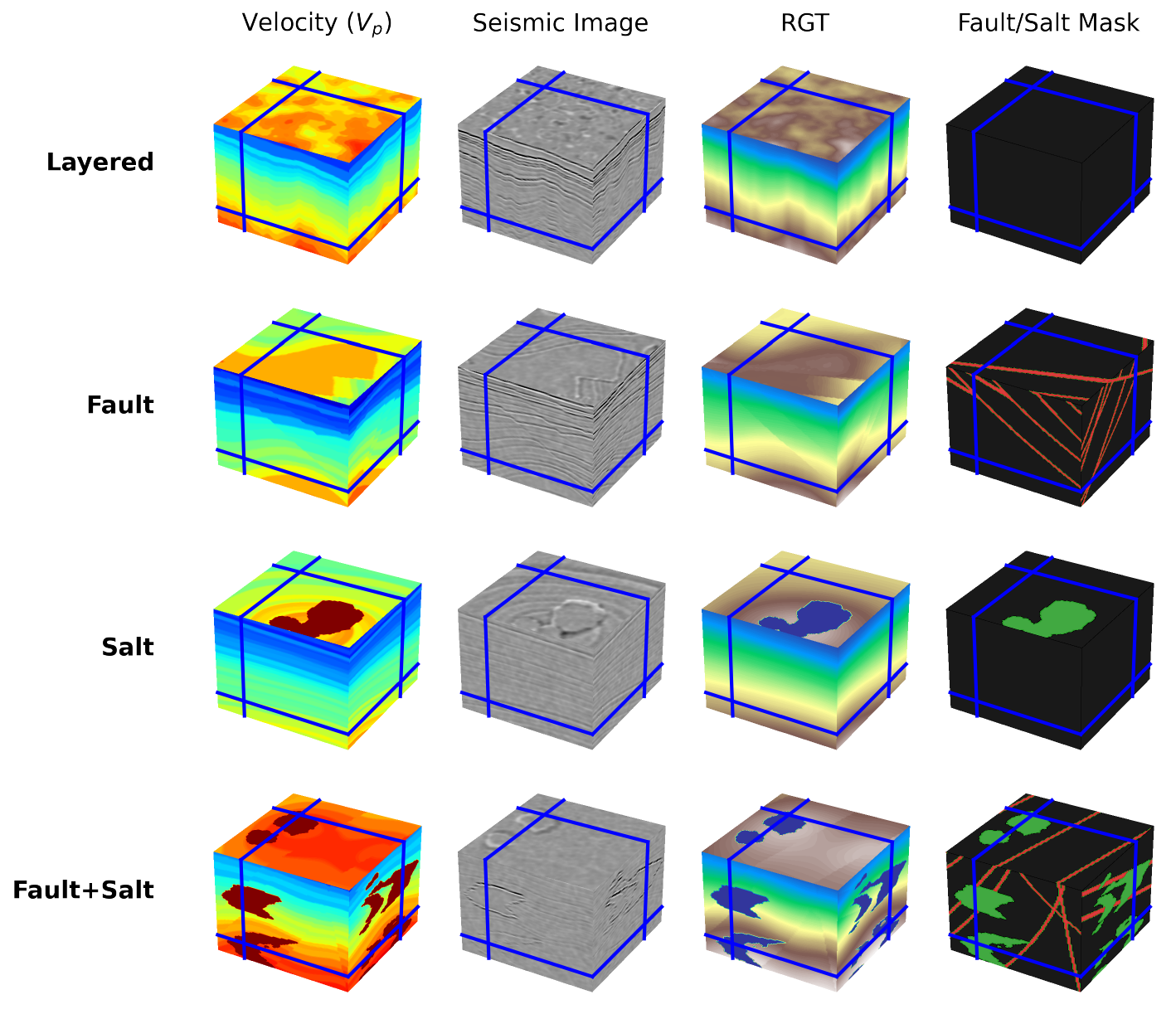}
\caption{Multi-modal 3D gallery highlighting representative volumetric realizations from each of the four structural complexity classes within the GeoFWI3D dataset. Rows (top to bottom) display examples of pure stratigraphy, faulted networks, salt diapirism, and complex coupled fault-salt models. Columns demonstrate the co-registered, dense labeling architecture of the repository: (column 1) compressional wave velocity ($V_p$) (column 2) zero-offset acoustic reflectivity volumes (column 3) continuous relative geologic time (RGT) field, illustrating the unphysical null-space (dark blue) generated by salt intrusion and (column 4) the combined discrete semantic mask, simultaneously tracking tectonic fault planes (red) and salt bodies (green) against the stratigraphic background.}
\label{fig:models3d}
\end{figure}

\begin{figure}[htbp]
\centering
\includegraphics[width=1.0\textwidth]{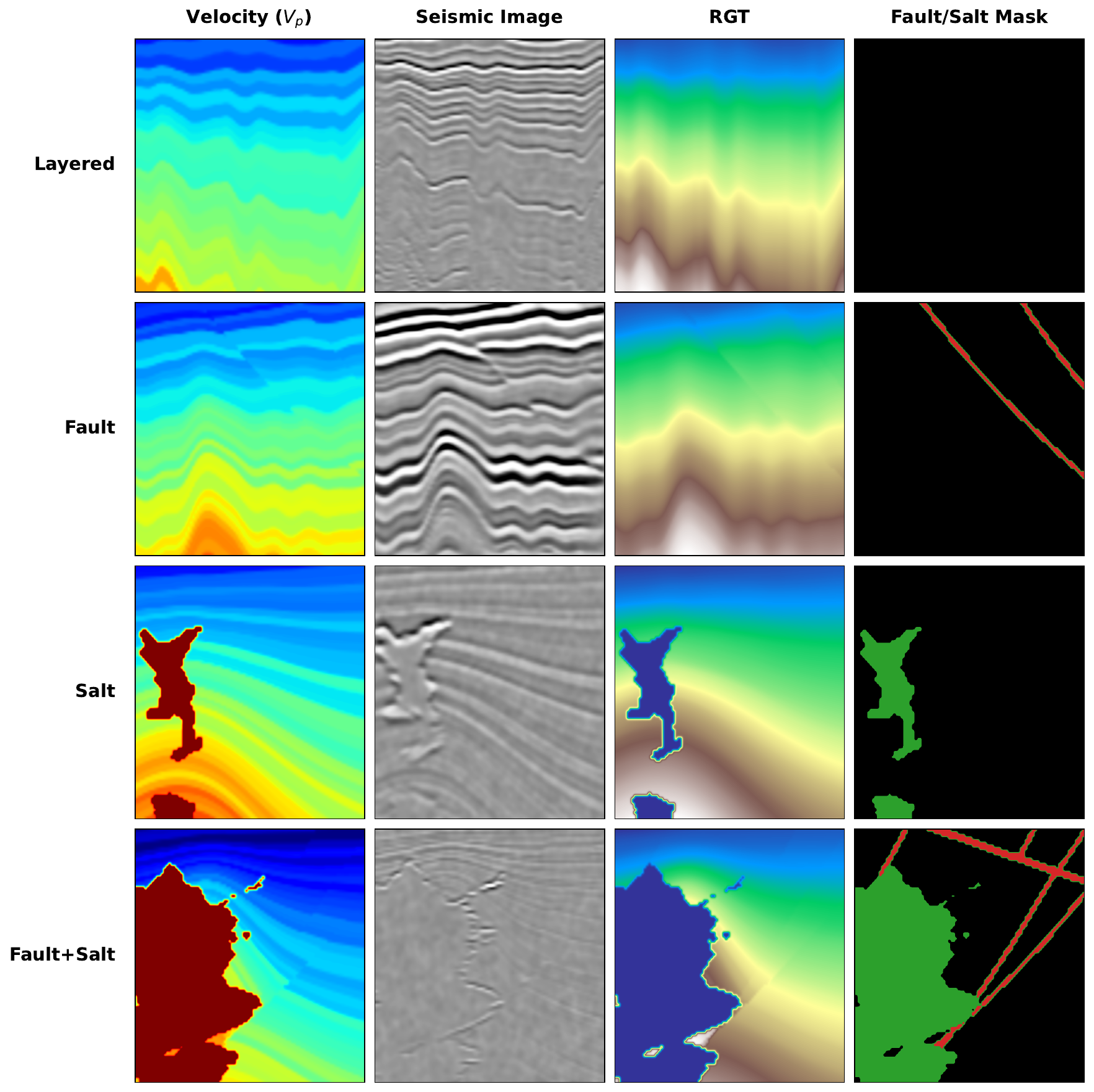} 
\caption{2D cross-sectional gallery illustrating the internal structural fidelity and co-registered multi-modal alignment of the GeoFWI3D dataset. Rows display vertical slices extracted from representative volumes across the four primary complexity classes: pure stratigraphy (layered), faulted networks, salt diapirism, and complex coupled deformations (fault + salt). Columns detail the distinct data modalities available for multi-task ML: (column 1) compressional wave velocity ($V_p$) profiles (column 2) corresponding acoustic reflectivity (column 3) continuous relative geologic time (RGT) fields, and (column 4) The unified semantic mask, labeling tectonic fault planes (red) and salt bodies (green) against the intact stratigraphic background (black).}
\label{fig:models2d}
\end{figure}

As mentioned earlier, the model generation strategy enforces a strict four-class structural-balancing protocol within the stochastic sampling loop. 
To visually demonstrate the structural diversity and multi-modal alignment of the GeoFWI3D datasets, representative 3D volumetric models from each of the four classes are presented in Figure \ref{fig:models3d}. The rows systematically progress through the dataset's structural hierarchy, beginning with purely depositional stratigraphy, followed by isolated faulted networks, then salt diapirism, and finally with complex, coupled deformations where intrusive salt bodies intersect tectonic faulting. Across the columns, the multi-model volumes are illustrated. The compressional velocity ($V_p$) serves as the petrophysical baseline that dictates the acoustic response of the synthetic seismic amplitude volumes. The corresponding label space is defined by the continuous RGT fields and the unified discrete semantic masks, which isolate both fault planes and salt boundaries from the background stratigraphy.
To further emphasize the internal structural fidelity of these structures, corresponding 2D inline slices are extracted and shown in Figure \ref{fig:models2d}. While the 3D gallery establishes the dataset's macroscopic geometric diversity, these 2D cross-sections demonstrate adherence to boundaries at the voxel level.  Table 1 describes the range of parameter space used for generating the full dataset.

\begin{table}[htbp]
\centering
\caption{Selected hyperparameter bounds for Latin Hypercube Sampling. Spatial parameters measured in grid points are scaled specifically for the $96 \times 96 \times 96$ computational volume.}
\label{tab:params}
\begin{tabular}{@{}llc@{}}
\toprule
\textbf{Category} & \textbf{Parameter Description} & \textbf{Sampling Bounds} \\ \midrule
\multirow{4}{*}{Stratigraphy} & Number of layers ($n_l$) (count) & [20, 40] \\
 & Vertical thickness variation ($l_{wv}$) (fraction) & [0.3, 0.5] \\
 & Horizontal thickness variation ($l_{wh}$) (fraction) & [0.2, 0.4] \\
 & Reflector height maximum (grid points) & [30.0, 70.0] \\ \midrule
\multirow{4}{*}{Fault Networks} & Number of faults ($n_f$) (count) & [4, 12] \\
 & Displacement range (grid points) & [3.0, 20.0] \\
 & Listric delta dip ($^\circ$) & [0.0, 40.0] \\
 & Conjugate strike spread ($^\circ$) & [80.0, 100.0] \\ \midrule
\multirow{5}{*}{Salt Bodies} & Number of salt bodies ($n_{salt}$) (count) & [1, 5] \\
 & Salt base radius (grid points) & [6.0, 21.0] \\
 & Top surface relative depth ($z$-fraction) & [0.4, 0.7] \\
 & Path lateral variation (grid points) & [5.0, 20.0] \\ 
 & Radius stochastic variation (fraction) & [0.3, 0.9] \\ \midrule
\multirow{3}{*}{Petrophysics} & Minimum background $V_p$ (m/s) & [1500, 2000] \\
 & Maximum background $V_p$ (m/s) & [3500, 4200] \\
 & Constant Salt $V_p$ (m/s) & [4480, 4530] \\ 
\bottomrule
\end{tabular}
\end{table}

\section{Data Records}

The complete GeoFWI3D dataset comprises 10,000 unique, geologically plausible 3D subsurface volumes designed to provide high-fidelity structural and petrophysical benchmarks for ML. 
The full dataset is hosted at \url{https://doi.org/10.5281/zenodo.20148778} \cite{swaminadhan_2026_20148778} and distributed under Creative Commons Attribution 4.0 International.

\subsection{Data Format and Storage Architecture}
To optimize network bandwidth and eliminate the local file system indexing overhead typical of large-scale DL repositories, the 10,000 models are batched and compressed into 10 individual tarball archives. Each archive contains a sequential block of 1,000 distinct models (e.g., \texttt{models\_batch\_0.tar.gz} containing models 0000 to 0999). 
All physical property volumes and structural labels are saved as raw binary arrays using a uniform 32-bit floating-point precision (\texttt{float32}) format. After extraction, the files automatically collapse into a standardized, flat directory hierarchy in which each geological realization is isolated by a unique, four-digit, zero-padded index ranging from \texttt{model\_0000} through \texttt{model\_9999}. Each individual subdirectory contains four co-registered 3D tensors:
\begin{itemize}
    \item \texttt{vp3d.bin}: The 3D compressional P-wave velocity model, with cell values stored explicitly in meters per second (m/s).
    \item \texttt{fault3d.bin}: The discrete 3D semantic mask tracking tectonic deformation, where a value of 0 indicates intact stratigraphy and non-zero integer values map specific, individual fault plane indices.
    \item \texttt{rgt3d.bin}: The continuous 3D Relative Geologic Time field, establishing a monotonically increasing coordinate space tied directly to depositional chronology.
    \item \texttt{image3d.bin}: The bandlimited zero-offset acoustic reflectivity volume, generated by convolving the broadband P-P reflectivity with a 3D spatial PSF in the wavenumber domain.
\end{itemize}
Binary masks for salt bodies can be efficiently generated on the fly by identifying voxels with zero values in the RGT volumes, since salt bodies physically overwrite and terminate the stratigraphic RGT field. Similarly, the background stratigraphic mask can be obtained by identifying all non-zero RGT coordinates. This implicit labeling approach reduces total dataset storage requirements.

Since these files are written as flat, headers-free binary streams, they can be directly mapped into computational memory objects (e.g., NumPy arrays or PyTorch tensors) using simple C-style array strides without requiring secondary file parsing, wrapper decoding, or custom metadata readers.

\subsection{Flexible Dataset Partitioning}
To maximize downstream utility across a broad spectrum of geophysical inversion tasks, the 10,000 volumes are intentionally delivered as a unified, unpartitioned repository rather than pre-split into training, validation, and test subsets. A comprehensive master metadata catalog (in a standard comma-separated format, \texttt{parameters.csv}) is included with the dataset. This gives researchers the complete operational flexibility to dynamically construct custom, stratified, or out-of-distribution (OOD) dataset splits tailored to specific ML experimental designs. The full dataset is uniformly divided into four classes. 2,500 volumes of pure stratigraphy (Class 1), 2,500 volumes of faulted networks (Class 2), 2,500 volumes of isolated salt diapirism (Class 3), and 2,500 volumes of complex coupled deformations (Class 4). Because the stochastic sampling uses a non-repeating, strictly sequential random-seeding protocol across the entire 10,000-volume pipeline, users can safely implement standard pseudo-random slicing, stratified k-fold cross-validation, or class-specific masking without risk of duplicate structural realizations.

\subsection{Comparison with Existing Benchmarks}
To highlight the unique structural contributions of the GeoFWI3D dataset, its features are compared with those of other publicly available 3D seismic ML datasets in Table 2. Most open-source data frameworks typically present a distinct trade-off between structural complexity and overall sample scale. 
GeoFWI3D addresses the lack of publicly available large-scale 3D benchmark datasets by delivering 10,000-volume 3D models that simultaneously incorporate multi-scale continuous stratigraphy, complex intersecting fault networks, and salt bodies, along with multi-modal semantic labels. This multi-modal labeling ensures that users do not have to choose between benchmarking inversion algorithms and structural interpretation networks, as both can be evaluated simultaneously on the same baseline subsurface models.

\begin{table}[htbp]
\centering
\caption{Comparative analysis of GeoFWI3D against existing 3D public seismic ML datasets across structural features, volume, and label variety.}
\label{tab:dataset_comparison}
\begin{threeparttable}
\resizebox{\textwidth}{!}{%
\begin{tabular}{@{} lcccccc @{}}
\toprule
\textbf{Dataset} & \textbf{Dimension} & \textbf{Faults} & \textbf{Salt} & \textbf{Dataset Size} & \textbf{Ground-Truth Labels} & \textbf{Open-Source} \\ \midrule
FaultSeg3D \cite{wu2019faultseg3d}     & 128 $\times$ 128 $\times$ 128                 & Yes         & No            & Small (220)               & Seismic reflectivity \& fault mask             & Yes                  \\
OpenFWI \tnote{*} \cite{deng2022openfwi}     & 350 $\times$ 400 $\times$ 400                 & No         & No            & Small (1827)               & Velocity \& Shot gathers            & Yes                  \\
\textbf{GeoFWI3D} & \textbf{96 $\times$ 96 $\times$ 96} &  \textbf{Yes}    & \textbf{Yes}  & \textbf{Large (10,000)} & \textbf{Multi-Modal Tensors\tnote{**}} & \textbf{Yes} \\ \bottomrule
\end{tabular}%
}
\begin{tablenotes}
\small
\item[*] OpenFWI is largely 2D models, we consider only the 3D models.
\item[**] Multi-modal tensors include structural velocity ($V_p$), continuous RGT, and discrete semantic fault \\ index masks, seismic volume.
\end{tablenotes}
\end{threeparttable}
\end{table}

\section{Technical Validation}

\subsection{Statistical characterization of geological parameter space}

To validate the stochastic integrity and fidelity of the GeoFWI3D dataset, we performed a comprehensive statistical analysis of the generated parameter space. The generation workflow is driven by the LHS to ensure a space-filling, near-random distribution that minimizes parameter correlation while maximizing the coverage of geological extremes. Ensuring flat, uniform distributions across these variables is critical as it helps prevent downstream neural networks from inadvertently learning biases toward specific thicknesses, velocities, or structures.

\begin{figure}[htbp]
\centering
\includegraphics[width=1.0\textwidth]{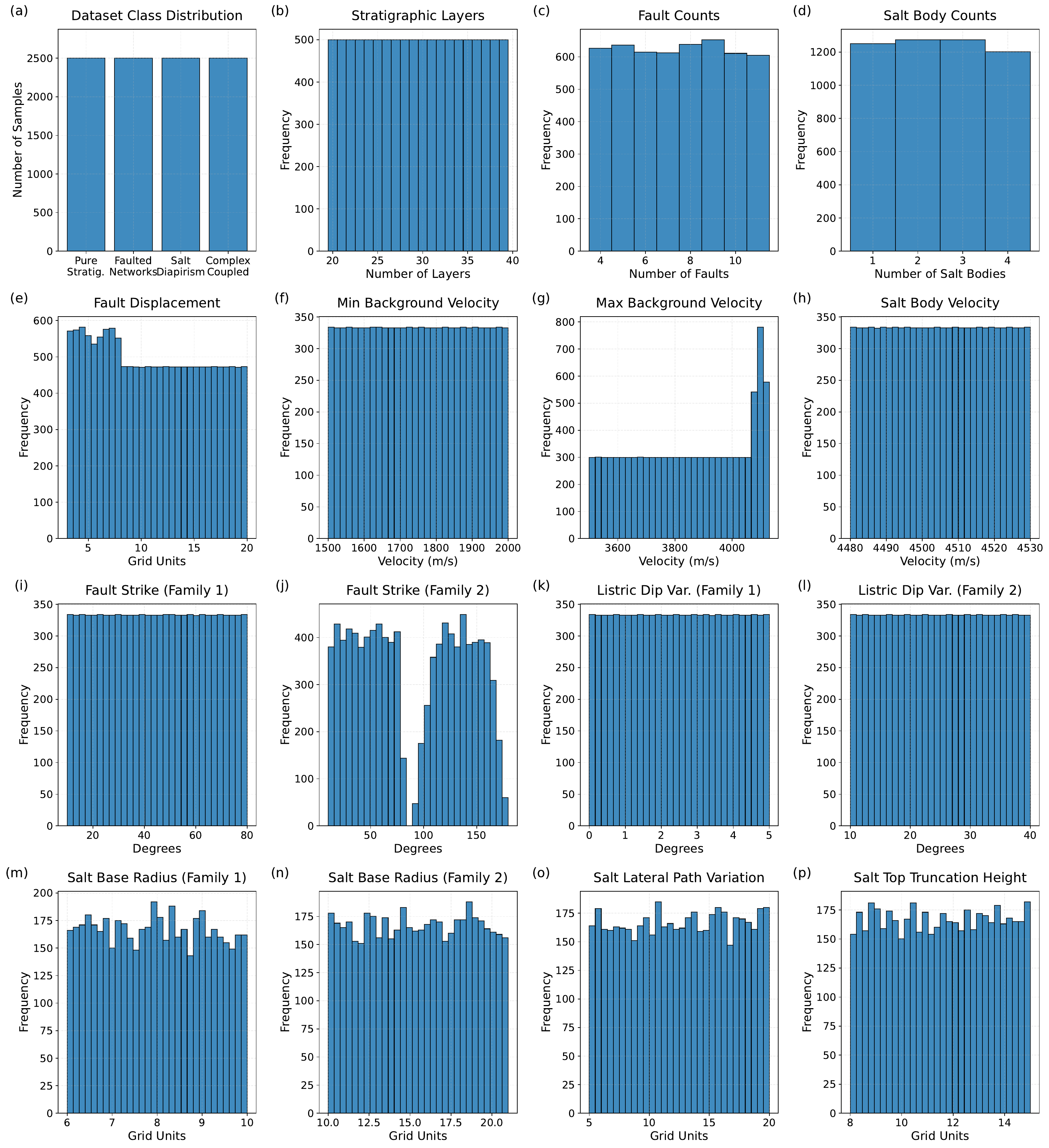} 
\caption{Statistical validation of the GeoFWI3D dataset parameter space. (a) Balanced distribution of the 10,000 samples across the four structural complexity tiers. (b–e) Frequency histograms for foundational structural elements, including stratigraphic layering, structural counts, and fault displacement magnitudes. (f–h) Petrophysical distributions validating the spread of minimum and maximum background $V_p$ along with targeted salt velocities. (i–l) Tectonic kinematic parameters demonstrating the uniform distribution of fault strike angles and listric curvature variations across fault families. (m–p) Geometric parameters controlling the dimensional diversity of salt bodies, including basal radii, lateral deviation, and truncation height.}
\label{fig:statistics}
\end{figure}

\noindent
\textbf{Class Balance and Structural Counts (Figures \ref{fig:statistics}a--e)}: The foundation of a robust ML benchmark is class equilibrium. As shown in Figure \ref{fig:statistics}a, the 10,000 models are partitioned into exactly 2,500 realizations for each of the four structural complexity tiers. This balance prevents DL models from class collapse during training. Beyond the categorical labels, the internal structural complexity is highly diverse. The number of stratigraphic layers (Figure \ref{fig:statistics}b) uniformly spans from 20 to 40, simulating varied depositional environments ranging from thick, massive beds to highly laminated, high-frequency sequences. Tectonic and intrusive events are similarly varied, with fault counts (Figure \ref{fig:statistics}c) and salt body counts (Figure \ref{fig:statistics}d) uniformly distributed to represent both sparse deformations and highly chaotic, highly fractured canopies. Fault displacement magnitudes (Figure \ref{fig:statistics}e) are uniformly sampled between 3 and 20 grid units, ensuring algorithms are trained on varying degrees of reflector discontinuity, which is challenging for automated horizon tracking.

\noindent
\textbf{Petrophysical Velocity Bounds (Figures \ref{fig:statistics}f--h)} \\
The acoustic impedance contrasts that drive seismic reflections are governed by the assigned compressional velocity ($V_p$). Minimum background velocities (Figure \ref{fig:statistics}f) are uniformly distributed between 1,500 and 2,000 m/s, representing the transition from the water column or unconsolidated near-surface sediments to shallow strata. Maximum background velocities (Figure \ref{fig:statistics}g) span up to 4,200 m/s to represent highly compacted, deep-basin lithologies. In contrast to the wide background ranges, the salt body velocities (Figure \ref{fig:statistics}h) are tightly constrained with a uniform distribution between 4,480 and 4,530 m/s. This narrower range is intentional, reflecting the geology of pure halite and evaporite intrusions compared to the surrounding clastic rocks. 

\noindent
\textbf{Tectonic Kinematics (Figures \ref{fig:statistics}i--l)} \\
To ensure tectonic realism, the dataset generates two distinct fault families capable of forming complex conjugate intersecting networks.
The primary fault strike angles (Figure \ref{fig:statistics}i) are sampled uniformly. However, to guarantee the formation of structurally complex, intersecting conjugate networks rather than parallel swarms, the secondary fault strikes (Figure \ref{fig:statistics}j) are generated with a significant angular offset from the primary family. This conditional sampling manifests as a bimodal uniform distribution, leaving a rotational gap to prevent co-linear structural alignment.
Real-world faults are rarely perfectly planar, they exhibit listric curvature, flattening at depth. The listric dip variation parameters (Figures \ref{fig:statistics}k, l) dictate this curvature. Family 1 represents planar to sub-planar faults with minor variation (0 to 5 degrees), while family 2 represents highly listric, major detachment faults with severe curvature (10 to 40 degrees).

\noindent
\textbf{Salt Geometries (Figures \ref{fig:statistics}m--p)} \\
Finally, the dimensional diversity of the salt bodies is statistically validated. Salt base radii are independently sampled across two families (Figures \ref{fig:statistics}m, n) to ensure a mix of narrow salt stocks and massive, broad diapirs within the same volume. The lateral path variation parameter (Figure \ref{fig:statistics}o) is critical as it is uniformly sampled up to 20 grid units. This parameter forces the salt bodies to deviate from a vertical ascent. This sinuosity creates structural overhangs and rugose boundaries, creating the root cause of illumination gaps in traditional FWI. Finally, the top truncation height (Figure \ref{fig:statistics}p) dictates the extrusion depth, ensuring a uniform mix of deep-seated and shallow surface-breaching salt domes.

\begin{figure}[htbp]
    \centering
    \includegraphics[width=0.85\linewidth]{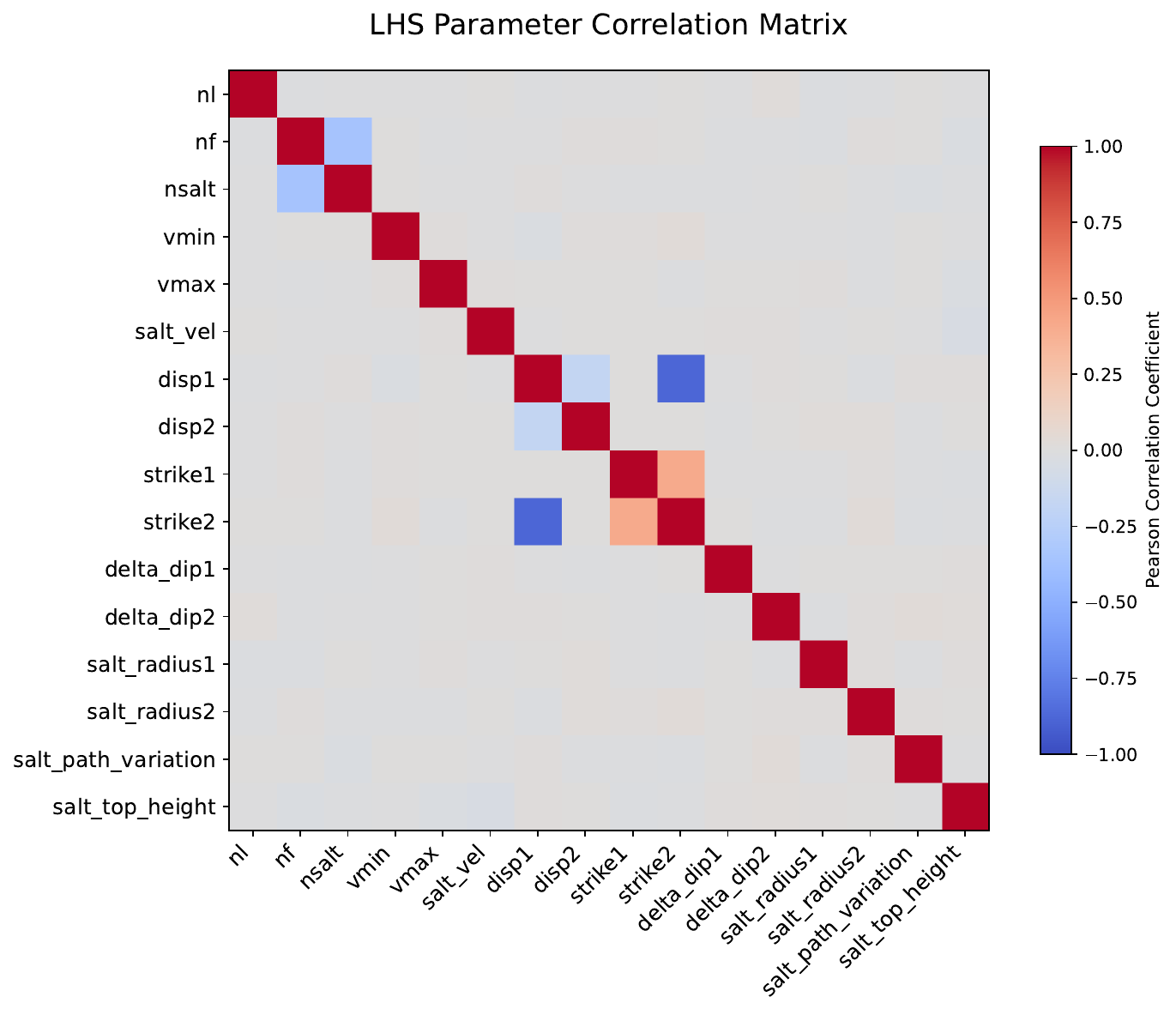} 
    \caption{Pearson correlation matrix of the Latin hypercube sampling parameter space. The near-zero off-diagonal values (in pale grey) verify the orthogonality of the sampled variables, ensuring that downstream neural networks learn the underlying geological topologies rather than statistical biases. The localized negative correlation between tectonic fault counts ($n_f$) and salt body counts ($n_{salt}$) is a deliberate, deterministic artifact of the dataset's strict categorical balancing protocol, which enforces mutually exclusive structural regimes in two of the four dataset classes.}
    \label{fig:corr-mat}
\end{figure}

To show that the structural diversity of GeoFWI3D prevents neural networks from spurious statistical biases, we evaluated the independence of the high-dimensional generation space. Figure \ref{fig:corr-mat} presents the Pearson correlation matrix for the primary LHS parameters. A robust ML benchmark requires near-zero correlation ($R \approx 0$) across independent physical variables. Otherwise, a network might improperly conflate distinct physical phenomena. 
The matrix demonstrates near-perfect orthogonality across the petrophysical and geometric variables, suggesting that the LHS protocol achieved a space-filling, unbiased distribution. 
The negative correlation observed between the number of faults ($n_f$) and the number of salt bodies ($n_{salt}$) is a deterministic consequence of the dataset's four-class categorical equilibrium. Because the generation workflow enforces mutually exclusive deformation regimes for Class 2 (faults present, salt explicitly bypassed) and Class 3 (salt present, faults explicitly bypassed), the counts for these two classes are distributed across 50\% of the dataset.
The localized correlations observed among the geometric parameters of the separate fault families (e.g., strikes and displacement magnitudes) arise from the geological rules governing conjugate fault networks. With intersecting conjugate regimes rather than parallel swarms, the framework couples secondary fault properties to the primary family. This conditional sampling promotes tectonic realism but inherently manifests as statistical correlation strictly within the fault parameter block.

\subsection{Fault detection}

To quantify the ML utility and out-of-distribution (OOD) generalization capacity of the GeoFWI3D repository, we established a comparative benchmark against FaultSeg3D \cite{wu2019faultseg3d}, a synthetic structural dataset. Because neural network performance is inextricably linked to both architecture and training data, we isolated the dataset's explicit contribution by training identical 3D U-Net architectures on both distributions with identical hyperparameters. 
Because seismic fault planes represent a negligible volumetric fraction of the entire 3D domain, standard loss formulations inherently bias the network toward the dominant background class. To mitigate this severe class imbalance, the network is optimized with a dynamically balanced cross-entropy loss. This formulation computes a scalar multiplier, $\beta$, that represents the ratio of non-fault to fault voxels in the true label, effectively weighting the cost of false negatives and forcing the network to prioritize structural lineaments. Further details regarding the network architecture and training strategy are described in Wu et al. \cite{wu2019faultseg3d}.

\begin{figure}
    \centering
    \includegraphics[width=0.99\linewidth]{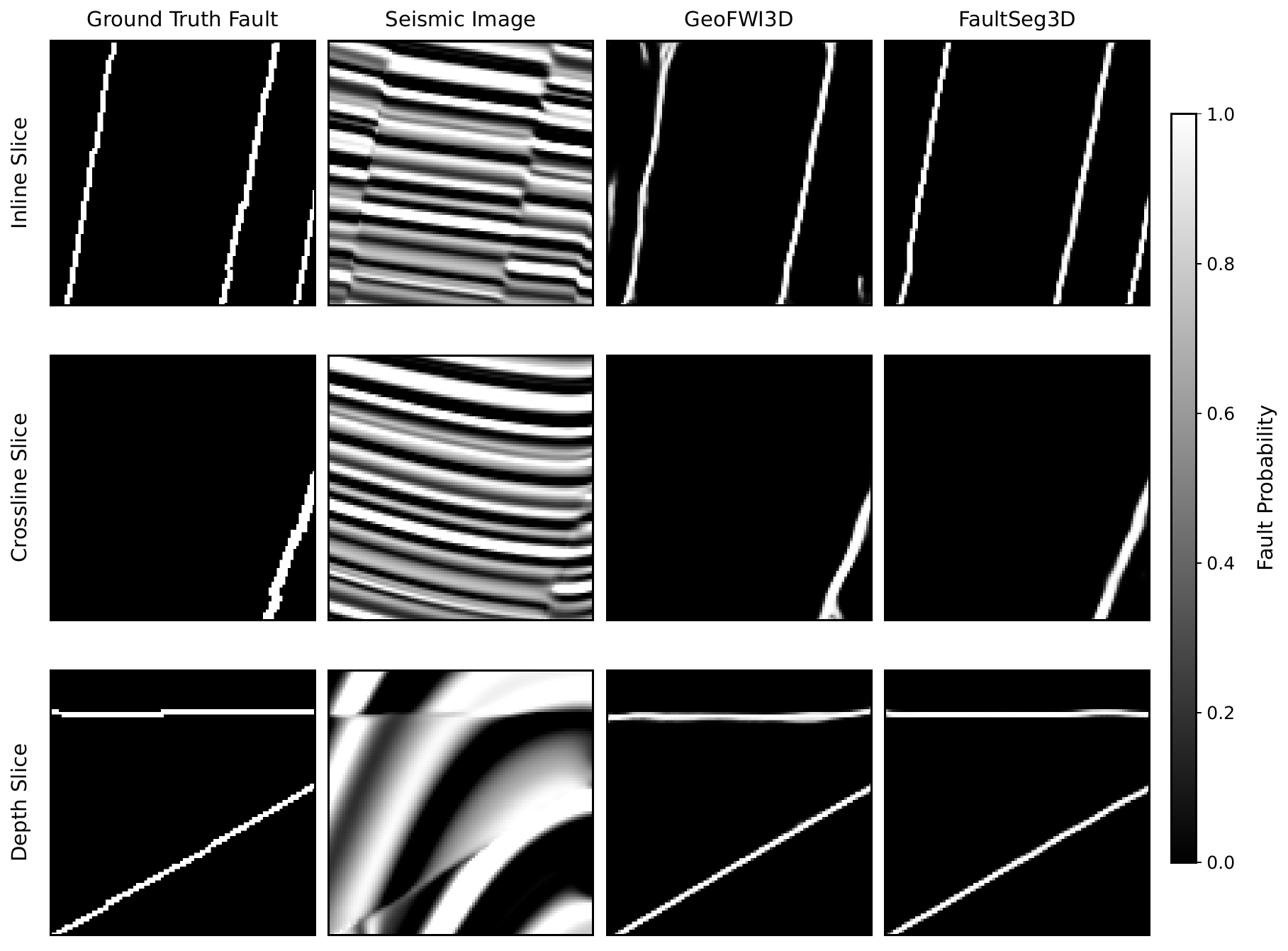}
    \caption{Visual comparison of fault detection performance on a representative volume from the baseline FaultSeg3D validation set. Rows display orthogonal views (inline, crossline, and depth slices) through the 3D volume. Columns correspond to the baseline ground-truth fault labels, the synthetic seismic amplitude, and the continuous probability predictions from networks trained on our GeoFWI3D dataset and the baseline FaultSeg3D dataset, respectively. Both models detect the dominant planar faults and are in close agreement, matching the ground truth label.}
    \label{fig:faultseg-validation}
\end{figure}

\begin{figure}
    \centering
    \includegraphics[width=0.99\linewidth]{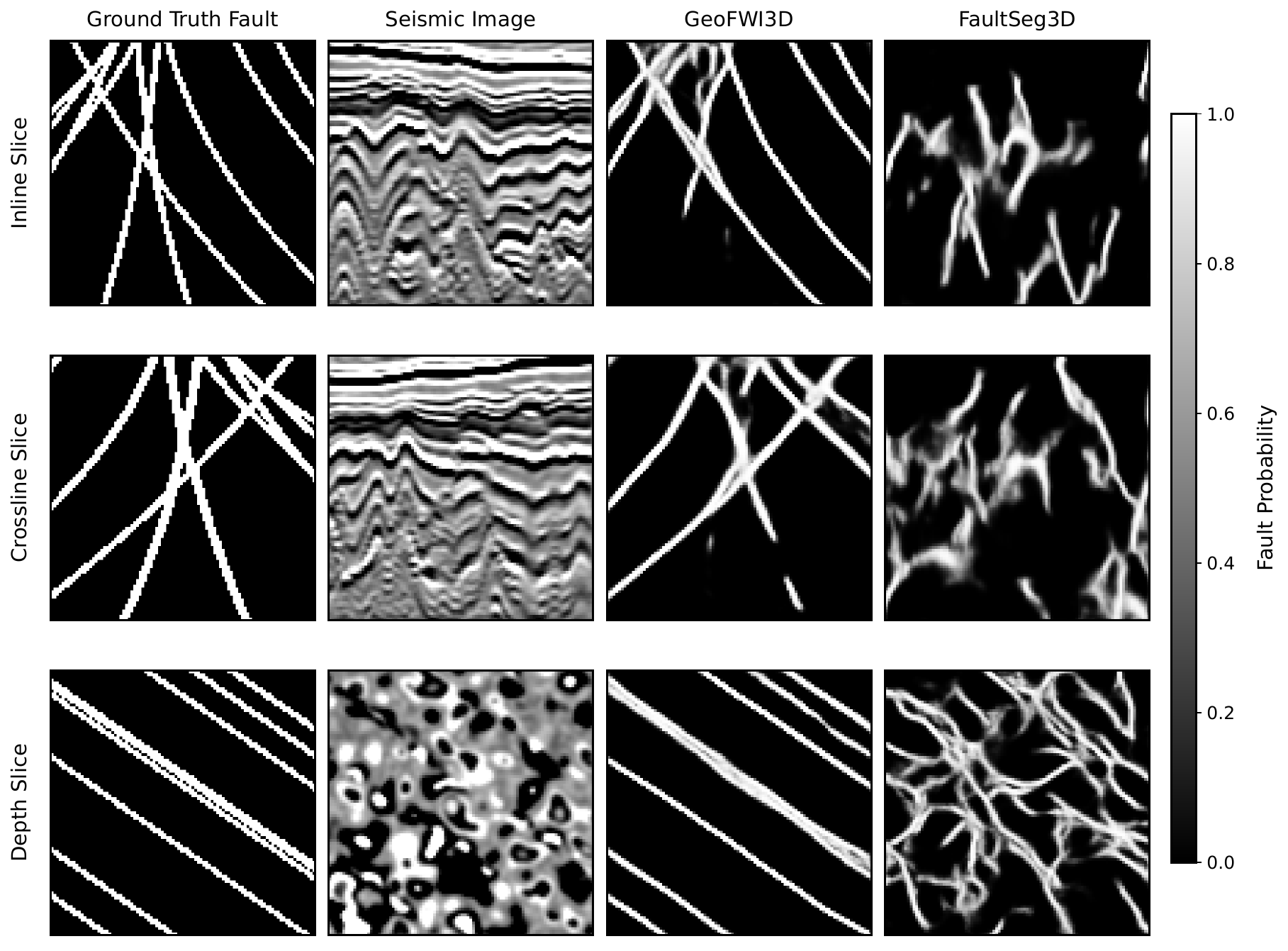}
    \caption{Visual comparison of fault detection performance on a complex, conjugate fault volume from the GeoFWI3D validation set. Rows display orthogonal views (inline, crossline, and depth slices) through the 3D volume. Columns correspond to the ground-truth fault labels, the synthetic seismic amplitude, and the continuous probability predictions from networks trained on the GeoFWI3D and baseline FaultSeg3D datasets, respectively. The GeoFWI3D-trained model resolves the intricate, cross-cutting tectonic networks with high fidelity and continuous slip planes. However, the FaultSeg3D baseline model gives incorrect predictions, unable to process intersecting geometries absent from its training distribution. It produces fragmented, disjointed probability fields and misses the dominant structural architecture.}
    \label{fig:geofwi-validation}
\end{figure}

To ensure fair comparison, the training sets for both networks were constrained to 200 samples, matching the maximum training samples of the legacy FaultSeg3D dataset. The training data was augmented systematically by
orthogonal flipping along the inline and crossline axes, and injecting random Gaussian noise.
The trained network was evaluated across three distinct domains: a native-distribution validation set (withheld during training), a reciprocal cross-domain validation set (evaluating each model on the opposing dataset), and a field application to the F3 field seismic volume from the North Sea. This multi-domain inference strategy is critical as it decouples a network's ability to overfit its native training distribution from its actual capacity to generalize to unseen, complex geological deformations. 
Figure \ref{fig:faultseg-validation} presents orthogonal cross-sections from the baseline FaultSeg3D validation set, comparing predictions from networks trained on both distributions. On this relatively simple and coherent seismic volume, both the GeoFWI3D and FaultSeg3D trained networks perform well, predicting the locations of the dominant planar faults in close agreement with the ground truth.
However, Figure \ref{fig:geofwi-validation} evaluates the models on the more structurally complex GeoFWI3D validation set, which contains dense intersecting and listric faults. On this volume, the GeoFWI3D-trained network resolves the complex cross-cutting geometries. The prediction from the FaultSeg3D-trained network, however, is incorrect and fails to accurately map the structural framework. 
This reduced cross-domain generalization occurs because the legacy FaultSeg3D training distribution lacks examples of intersecting geometries or conjugate stress regimes. Because of that, only a network exposed to these diverse, cross-cutting structural variations of the GeoFWI3D dataset during training can interpret such realistic structural complexity.

Table \ref{tab:cross_validation} summarizes the quantitative results of the cross-dataset evaluation conducted on 20 unseen volumes from both distributions. Both networks demonstrate strong baseline performance on their native test sets, achieving comparable F1-Scores of 0.7909 (FaultSeg3D) and 0.7769 (GeoFWI3D). 
However, evaluating the models out-of-distribution (OOD) reveals significant asymmetries in their ability to generalize. Notably, the network trained on the GeoFWI3D dataset performs better on the legacy FaultSeg3D validation set than the FaultSeg3D-trained network performs on the GeoFWI3D dataset. 
The GeoFWI3D-trained model achieves an OOD F1-Score of 0.4110 when evaluated on the FaultSeg3D dataset. In contrast, the FaultSeg3D-trained model suffers a performance degradation on the GeoFWI3D dataset, yielding an F1-Score of only 0.3023. This discrepancy indicates the incompleteness in the complexity of the legacy training distribution. 
While both models experience a metric decline when tested out-of-distribution, the GeoFWI3D-trained network remains more robust, retaining the structural capacity to interpret geology across varying levels of tectonic complexity.

\begin{table}[htbp]
\centering
\caption{Cross-dataset quantitative validation metrics. Models trained independently on GeoFWI3D and FaultSeg3D were evaluated against 20 unseen validation volumes from both their native and opposing distributions.}
\label{tab:cross_validation}
\begin{tabular}{llcccc}
\toprule
\textbf{Validation Set} & \textbf{Training Dataset} & \textbf{Precision} & \textbf{Recall} & \textbf{F1-Score} & \textbf{IoU} \\
\midrule
\multirow{2}{*}{FaultSeg3D} 
& FaultSeg3D (Native) & 0.7979 & 0.7840 & 0.7909 & 0.6541 \\
& GeoFWI3D (OOD)      & 0.4619 & 0.3701 & 0.4110 & 0.2586 \\
\midrule
\multirow{2}{*}{GeoFWI3D}   
& GeoFWI3D (Native)   & 0.7920 & 0.7624 & 0.7769 & 0.6352 \\
& FaultSeg3D (OOD)    & 0.4975 & 0.2171 & 0.3023 & 0.1781 \\
\bottomrule
\end{tabular}
\end{table}

\begin{figure}
    \centering
    \includegraphics[width=0.9\linewidth]{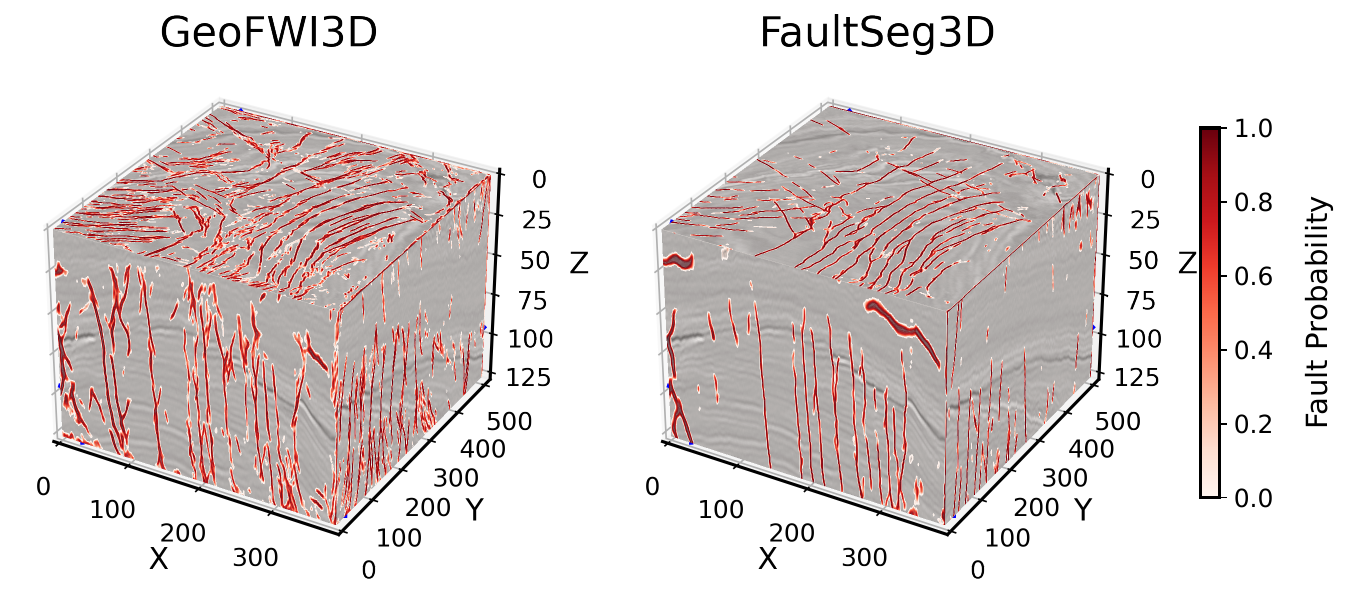}
    \caption{3D plots of fault detection on the F3 field seismic data from the North Sea, comparing the generalization capabilities of identical models trained on GeoFWI3D and FaultSeg3D datasets. The continuous probability fields (red) are overlaid on the orthogonal seismic amplitude slices. Left: The GeoFWI3D-trained network resolves a highly dense, spatially coherent 3D conjugate fault system, maintaining the continuity of slip planes across complex intersections. Right: The FaultSeg3D baseline model exhibits spatial fragmentation. While it detects isolated planar segments, it fails to map the connectivity of the network, resulting in false negatives and a loss of 3D structural integrity. This suggests that the complex parameter space of GeoFWI3D is valuable for training models capable of full-volume structural interpretation in the field.}
    \label{fig:f3-pred3d}
\end{figure}

While the cross-domain validation on synthetic holdout sets shows that the GeoFWI3D-trained network can resolve intersecting conjugate networks, synthetic data inherently represents a simplified approximation of the Earth. To illustrate the practical utility of the GeoFWI3D dataset, we deployed both the GeoFWI3D and FaultSeg3D trained networks on a 3D field seismic volume. Because field data lacks a definitive ground-truth label, this evaluation serves as a qualitative test of out-of-distribution (OOD) generalization under real-world acquisition noise, variable illumination, and true tectonic complexity.
As illustrated in the 3D volumetric plots (Figure \ref{fig:f3-pred3d}), the disparity in spatial coherence between the two models is striking. The baseline FaultSeg3D network produces a fragmented prediction. While it successfully flags isolated segments of major planar faults, it fails to connect them into continuous surfaces, leaving unmapped gaps within the conjugate swarm and producing low-resolution probability artifacts. 
On the other hand, the GeoFWI3D-trained network maps the complex tectonic architecture as a unified, spatially continuous 3D system. Because its training distribution explicitly modeled multiple fault families, varying listric curvatures, and dense intersecting networks, it tracks both dominant regional faults and subtle, cross-cutting conjugate fractures across the entire field volume. This suggests that the wide parameter space and structural heterogeneity of GeoFWI3D make it a more valuable dataset for training advanced architectures.

\subsection{Joint Multi-Modal Salt Segmentation and Continuous RGT mapping}

To demonstrate the multi-modal utility of the GeoFWI3D dataset, we formulated a simultaneous classification and regression workflow by developing a multi-task 3D convolutional architecture, denoted as SaltRGT3DNet, designed to learn a generalized latent representation of the subsurface from $64 \times 64 \times 64$ seismic patches.  Unlike a standard U-Net, SaltRGT3DNet operates as a bifurcated ``Y-Net," utilizing a shared encoder-bottleneck structure that splits into two task-specific decoders. To handle complex structural topologies, the network integrates 3D Dense blocks \cite{huang2017densely} to maximize feature reuse, paired with Selective Kernel (SK) attention modules \cite{li2019selective} that dynamically scale spatial receptive fields.
The network was trained by simultaneously minimizing mean squared error (MSE) loss for continuous RGT prediction and binary cross-entropy (BCE) loss for discrete salt segmentation.

\begin{figure}[htbp]
    \centering
    \includegraphics[width=0.9\linewidth]{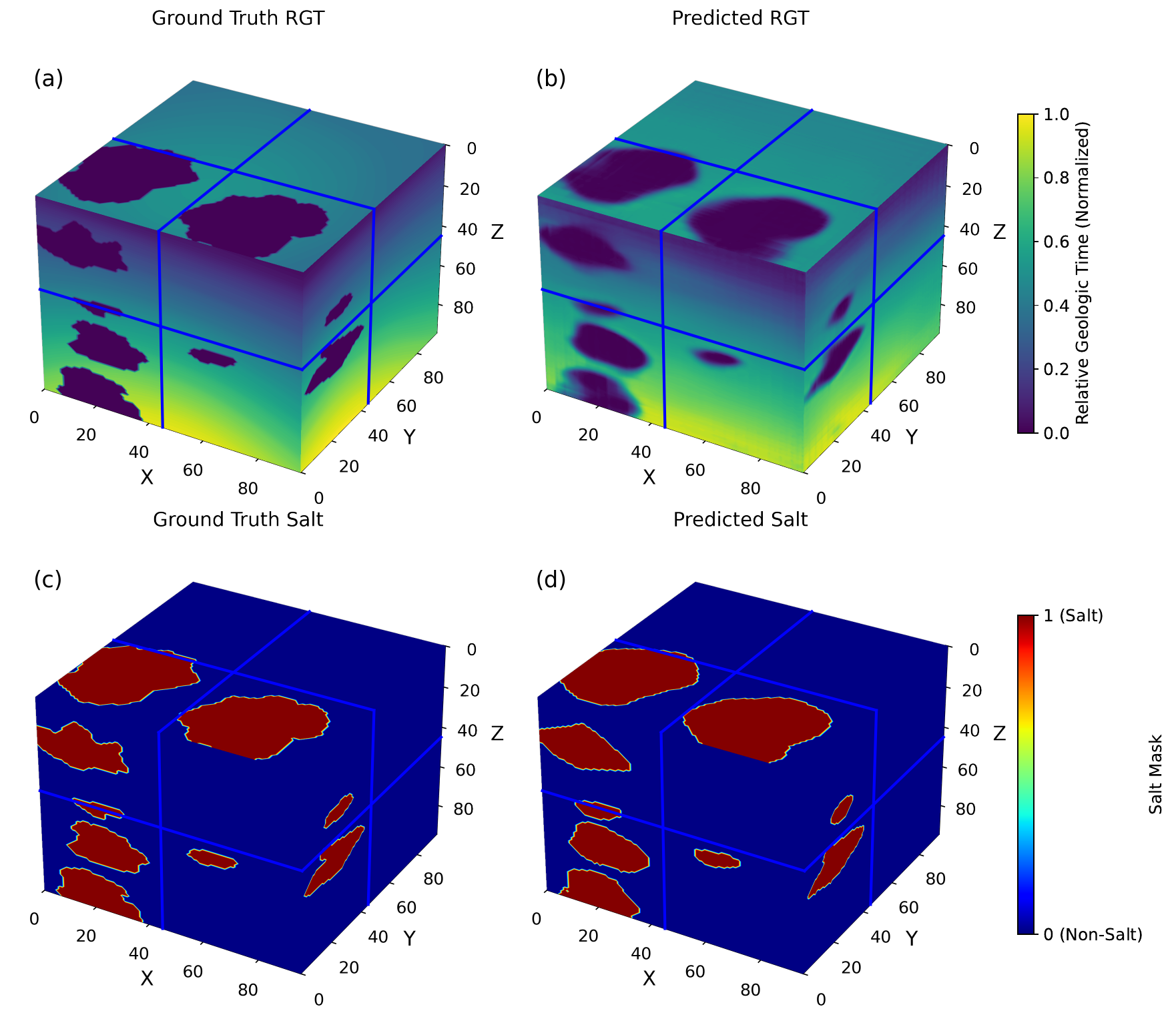}
    \caption{Joint multi-modal prediction on an unseen synthetic test volume from the GeoFWI3D dataset. (a) Ground truth relative geologic time (RGT) and (b) the corresponding continuous RGT prediction, demonstrating the network's ability to track stratigraphic horizons. (c) Ground truth binary salt mask and (d) the predicted salt mask, delineating the sharp boundary of the salt bodies.}
    \label{fig:rgt-salt-pred}
\end{figure}

The trained network evaluated on an unseen synthetic volume from the GeoFWI3D test set exhibits promising baseline performance. Concurrently, the continuous RGT regression branch tracks seismic horizons across the 3D domain (Figures \ref{fig:rgt-salt-pred}a, b), yielding a mean absolute error (MAE) of 0.0539 and an $R^2$ coefficient of 0.8995. Notably, when evaluating the RGT prediction exclusively in the non-salt, stratigraphically coherent regions of the volume, the performance increases to an $R^2$ of 0.9081 and an MAE of 0.0505. This suggests that the network resolves the underlying depositional kinematics without being affected by the amplitude shadowing of massive salt diapirs.

\begin{figure}[htbp]
    \centering
    \includegraphics[width=.99\linewidth]{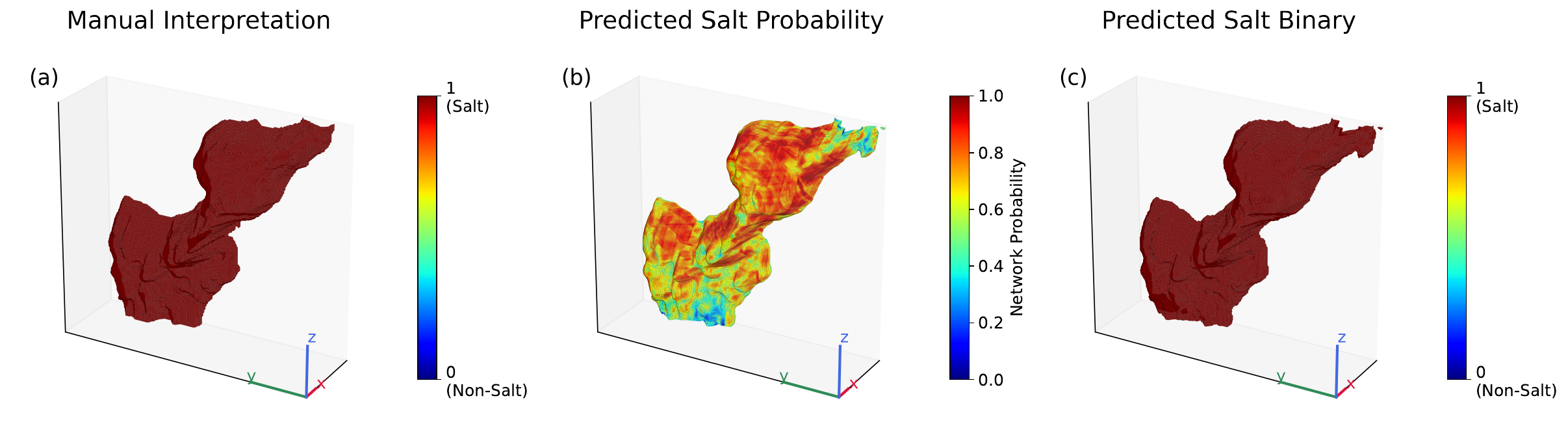}
    \caption{3D visualization of the salt body prediction. (a) 3D salt body obtained from manual interpretation (b) The continuous salt probability prediction obtained from network prediction (c) The final binary salt segmentation mask, confirming the model's ability to delineate sharp boundaries from the surrounding stratigraphy.}
    \label{fig:f3-salt-pred-marching-cubes}
\end{figure}

To demonstrate the field applicability of the GeoFWI3D synthetic pretraining, the synthetic-trained SaltRGT3DNet was adapted for a few-shot transfer learning task on the F3 field seismic volume. To simulate a limited-label scenario, the network's shared encoder was completely frozen to retain the foundational structural filters learned natively from GeoFWI3D. Only the bottleneck and salt decoder were fine-tuned, utilizing only 10\% of the available F3 data patches. The fine-tuned model evaluated on the remaining 90\% of the unseen real F3 data demonstrates encouraging generalization performance. The network successfully isolates the massive, complex salt geometry present in the seismic volume, achieving an accuracy of 0.9947, a precision of 0.9753, and an F1-score of 0.9782 (Figure \ref{fig:f3-salt-pred-marching-cubes}). This volumetric segmentation indicates that the complex, multi-modal features learned within the synthetic GeoFWI3D dataset distribution translate into effective structural priors for real-world seismic interpretation.

\subsection{Full Waveform Inversion}

\begin{figure}[htbp]
    \centering
    \includegraphics[width=0.99\linewidth]{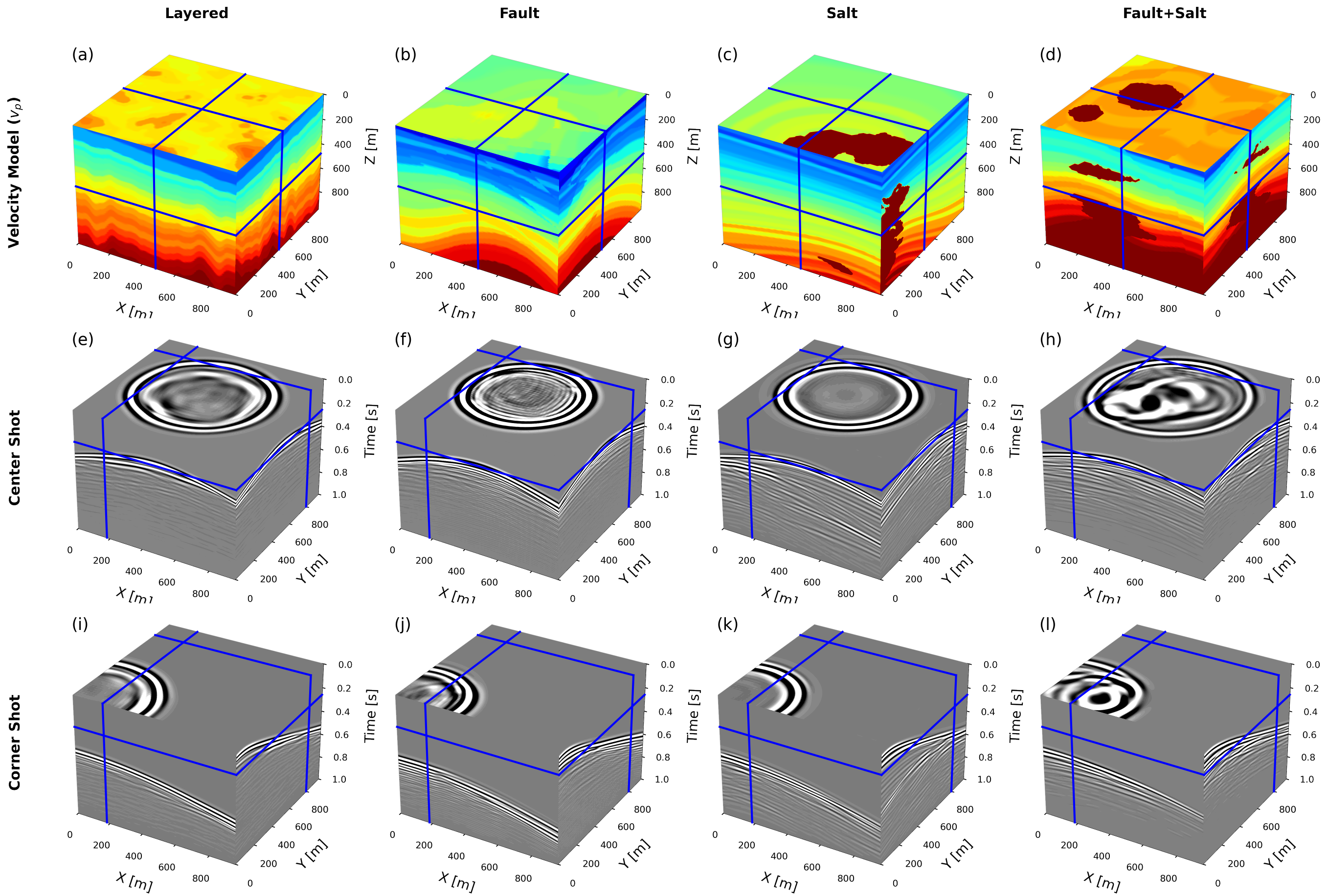}
    \caption{Representative 3D velocity models and their corresponding synthetic acoustic wavefields across the four structural classes of the GeoFWI3D dataset. The top row (a–d) displays the velocity ($v_p$) models with varying levels of complexity. The middle row (e–h) and bottom row (i–l) illustrate the full 3D time-domain wavefield responses originating from center and corner surface shot locations, respectively. The gathers exhibit severe scattering, diffractions, and phase distortions as structural complexity increases, providing the rich data variance necessary for robust, data-driven FWI.}
    \label{fig:shot-gathers}
\end{figure}

The raw GeoFWI3D models are stored on dimensionless computational grids. To demonstrate their physical validity and direct applicability to wave-equation-based workflows, we implemented a conventional 3D acoustic FWI. We used a spatial discretization of 10 m in all directions ($dx = dy = dz = 10$~m), yielding a physical volumetric domain of $960 \times 960 \times 960$~m. 
Before executing the inversion, we evaluated the forward acoustic wavefield responses across different classes. 
Figure \ref{fig:shot-gathers} illustrates representative 3D velocity models from each of these four classes alongside their corresponding acoustic wavefield responses with finite difference modeling using deepwave \cite{richardson_alan_2023}. It uses a fourth-order centered finite difference in space and a second-order discretization in time. Free surface boundary is incorporated at the top boundary to mimic air-water interface, and perfectly matching layer (PML) boundaries are added on all other boundaries to suppress boundary reflections \cite{berenger1994perfectly}.
By simulating full 3D shot gathers from source positions at the model's center and corners, the dataset captures the spectrum of wave propagation. In layered models, the wavefronts remain relatively coherent and continuous. However, as discrete structural heterogeneities are introduced, the wavefields exhibit severe kinematic distortions. The sharp velocity contrasts of the fault planes generate diffraction tails, while the rugose boundaries and high-velocity roots of the salt bodies induce wavefield scattering, multi-pathing, and illumination shadows. This explicit structural progression represents that the training dataset provides the data diversity required for ML applications.

\begin{figure}[htbp]
    \centering
    \includegraphics[width=.99\linewidth]{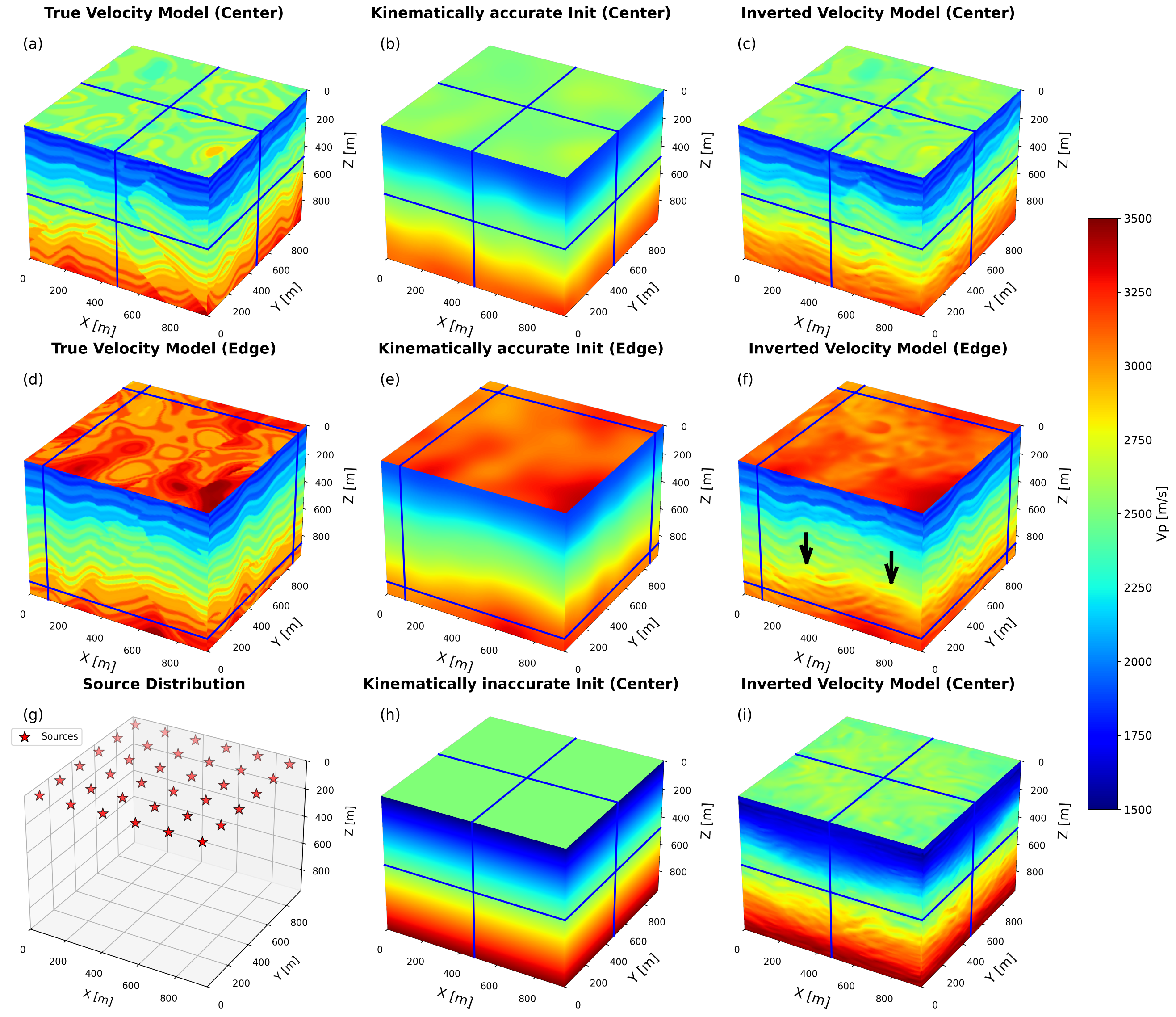}
    \caption{Baseline 3D structural velocity recovery and cycle-skipping using conventional gradient-based acoustic FWI. (a) The true 3D velocity model, featuring high-wavenumber folding and tectonic faulting. (b) A smoothed, kinematically accurate initial velocity model. (c) The inverted velocity model. (d-f) orthogonal slices corresponding to the edge of the model. (g) Spatial distribution of sources. (h) A kinematically inaccurate initial model defined by a simple 1D vertical depth gradient, which lacks the low-frequency macroscopic trend of the true stratigraphy and (i) the inversion result.}
    \label{fig:model0000-fwi}
\end{figure}

While the core GeoFWI3D dataset natively provides zero-offset reflectivity volumes, FWI requires raw pre-stack data. Therefore, we simulated synthetic 3D shot gathers using a 25 Hz dominant-frequency Ricker wavelet. The acquisition geometry consisted of 36 sources deployed in a $6 \times 6$ areal grid at a depth of 10 m, with a horizontal spacing of 170 m. The full 3D wavefield was recorded by a dense surface receiver array comprising a $96 \times 96$ grid, placing a receiver at every surface grid node. 
The inversion was driven by a standard multi-scale strategy, sequentially expanding the inversion bandwidth with upper-frequency cutoffs of 10, 15, and 25 Hz to mitigate cycle-skipping. Figure \ref{fig:model0000-fwi} illustrates the true velocity model, the  smooth initial model, and the final high-resolution inverted volume. The inversion posed as least square data misfit minimization, reduced the model MAE from 70 m/s to 49 m/s. The baseline results are promising as the conventional gradient-based scheme successfully updates the smooth kinematic background to recover high-wavenumber structural features, resolving both the major stratigraphic layering and the sharp fault lineaments. 

The spatial distribution of sources is illustrated in Figure \ref{fig:model0000-fwi}g. This acquisition geometry provides sufficient aperture to illuminate the core of the target structure. The orthogonal cross-sections displayed in Figure \ref{fig:model0000-fwi}a--c intersect the center of the 3D domain, where multi-azimuth ray-path coverage is optimal. Because of this dense central illumination, the inverted models exhibit the strongest agreement with the true velocity model within these interior regions. However, wavefield energy diverges and degrades toward the lateral boundaries. While the 36-source configuration adequately resolves the central macro-model, the deeper sections near the edges suffer from sparse coverage and these regions exhibit lower resolution updates marked by black arrows (Figure \ref{fig:model0000-fwi}d--f).

However, the success of conventional FWI remains dependent on the quality of the starting model. To demonstrate the severe non-linearity and ill-posed nature of this optimization, we performed a parallel inversion on the same volume using a kinematically inaccurate initial model consisting of a simple 1D vertical velocity gradient (Figure \ref{fig:model0000-fwi}h). As shown in Figure \ref{fig:model0000-fwi}i, this inversion struggles to reconstruct the geological structures. Because the starting model lacks the macroscopic structural trend, the phase difference between the modeled and observed wavefields exceeds a half-wavelength. This causes the optimizer to suffer from cycle-skipping, trapping the inversion in a local minimum and generating artifacts.

\begin{figure}[htbp]
    \centering
    \includegraphics[width=.99\linewidth]{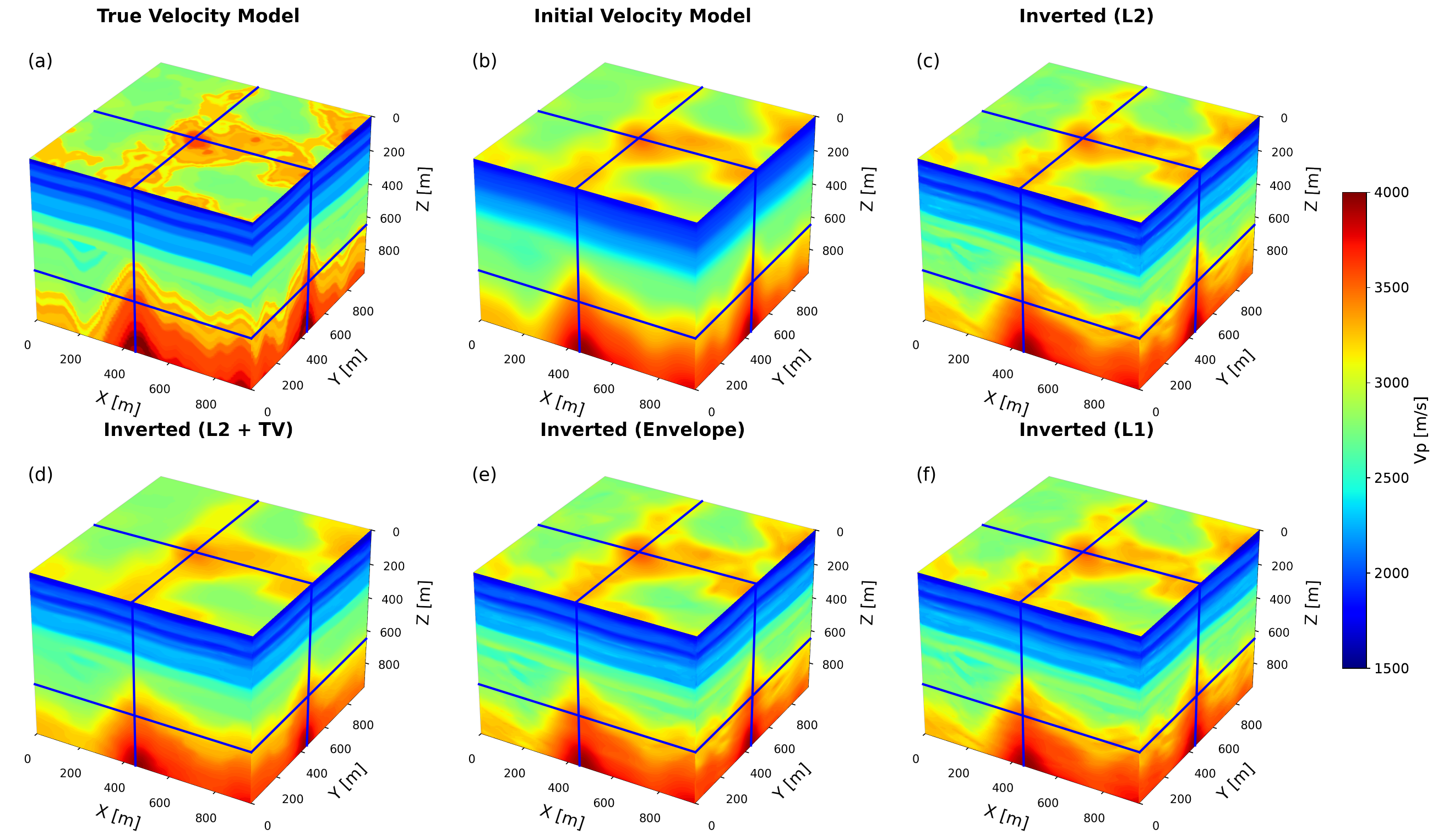}
    \caption{Comparison of velocity models utilizing different objective functions. 
    (a) True velocity model. 
    (b) Smooth initial velocity model. 
    (c) Inversion utilizing standard $L_2$ loss. 
    (d) Inversion utilizing $L_2$ loss coupled with Total Variation (TV) regularization. 
    (e) Inversion utilizing an envelope-based objective function. 
    (f) Inversion utilizing standard $L_1$ loss. 
    The $L_1$ formulation achieves the highest fidelity in recovering both the macro-model trends and the high-wavenumber structural boundaries.}
    \label{fig:model9999-fwi}
\end{figure}

To quantify the impact of the objective function on convergence, the FWI was executed using four distinct loss functions starting from the same smoothed initial model (Figure \ref{fig:model9999-fwi}b). The quantitative performance of each inversion was evaluated using MAE and root mean square error (RMSE) against the true velocity model. The initial model exhibited an MAE of 60.25~m/s and an RMSE of 74.07~m/s.
The standard point-by-point waveform difference metrics achieved the most accurate final velocity models, indicating that the initial model possessed sufficient kinematic accuracy to avoid cycle-skipping. The $L_1$ norm yielded the lowest quantitative errors across all tests (MAE: 43.00~m/s, RMSE: 58.46~m/s), followed closely by the $L_2$ norm (MAE: 44.13~m/s, RMSE: 59.81~m/s). Visually, these inversions (Figure \ref{fig:model9999-fwi}f, c) successfully recovered high-wavenumber details and sharpened the deep, high-contrast boundaries. The efficiency of the $L_1$ norm suggests it is more robust to the large, localized amplitude residuals generated by the high-velocity structures at the base of the model, whereas the $L_2$ norm's squared penalty may overly smooth these sharp interfaces.
The inversion minimizing envelope misfit (Figure \ref{fig:model9999-fwi}e) successfully updated the macroscopic velocity trends, reducing the errors to an MAE of 47.42~m/s and an RMSE of 63.48~m/s. Because the envelope objective acts as a low-pass filter by minimizing the analytic signal magnitude rather than the phase, it remains robust but lacks the high-frequency sensitivity required to recover sharp structural interfaces to the same degree as the $L_1$ or $L_2$ formulations.
The application of Total Variation (TV) regularization \cite{rudin1992nonlinear} to the $L_2$ objective ($L_2$ + TV) resulted in an MAE of 49.90~m/s and an RMSE of 67.42~m/s. While this formulation improved upon the initial model, it underperformed compared with the unregularized $L_1$ and $L_2$ objectives. As observed in Figure \ref{fig:model9999-fwi}d, the TV penalty forces a piecewise-constant, blocky structure. In a geological setting characterized by continuous velocity gradients alongside complex structural folding, this regularization over-constrains the gradient, preventing the optimizer from fully resolving the natural continuous variations present in the true model.

\begin{figure}
    \centering
    \includegraphics[width=.99\linewidth]{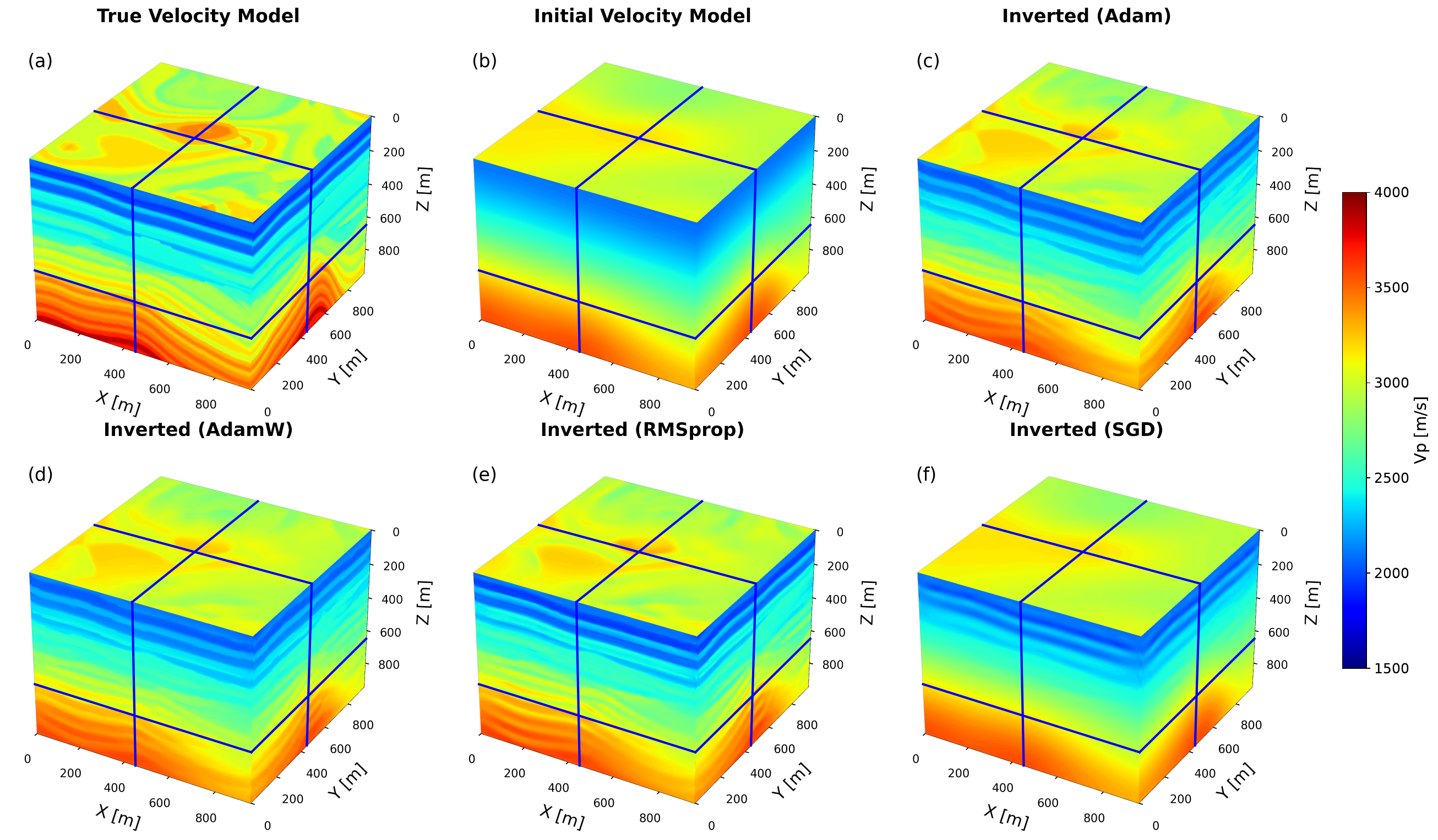}
    \caption{Comparison of velocity models utilizing different optimization algorithms. 
    (a) True velocity model. 
    (b) Smooth initial velocity model. 
    (c) Inversion utilizing the Adam optimizer. 
    (d) Inversion utilizing the AdamW optimizer. 
    (e) Inversion utilizing the RMSprop optimizer. 
    (f) Inversion utilizing Stochastic Gradient Descent (SGD). 
    RMSprop yields the highest structural resolution and lowest quantitative error, recovering sharp stratigraphic layering that the other optimizers fail to resolve.}
    \label{fig:optimizers-fwi}
\end{figure}

To evaluate convergence behavior in FWI, we compared four gradient-based optimization algorithms: Adam, AdamW, RMSprop, and SGD, all initialized from the same smoothed starting model (Figure \ref{fig:optimizers-fwi}b). RMSprop achieved the best performance, yielding the lowest errors (MAE: 64.28~m/s, RMSE: 87.13~m/s) and recovering sharper deep structures and higher-wavenumber stratigraphic features (Figure \ref{fig:optimizers-fwi}e). Adam and AdamW produced comparable results (MAE $\sim$81.28~m/s, RMSE $\sim$101.22~m/s), adequately reconstructing large-scale velocity trends (Figure \ref{fig:optimizers-fwi}c,d). SGD showed the weakest convergence, with minimal improvement over the initial model (MAE: 88.37~m/s, RMSE: 106.95~m/s (Figure \ref{fig:optimizers-fwi}f). These results indicate that adaptive learning-rate methods, particularly RMSprop, are effective for addressing the non-convex optimization landscape of FWI.

\begin{figure}[htbp]
    \centering
    \includegraphics[width=.99\linewidth]{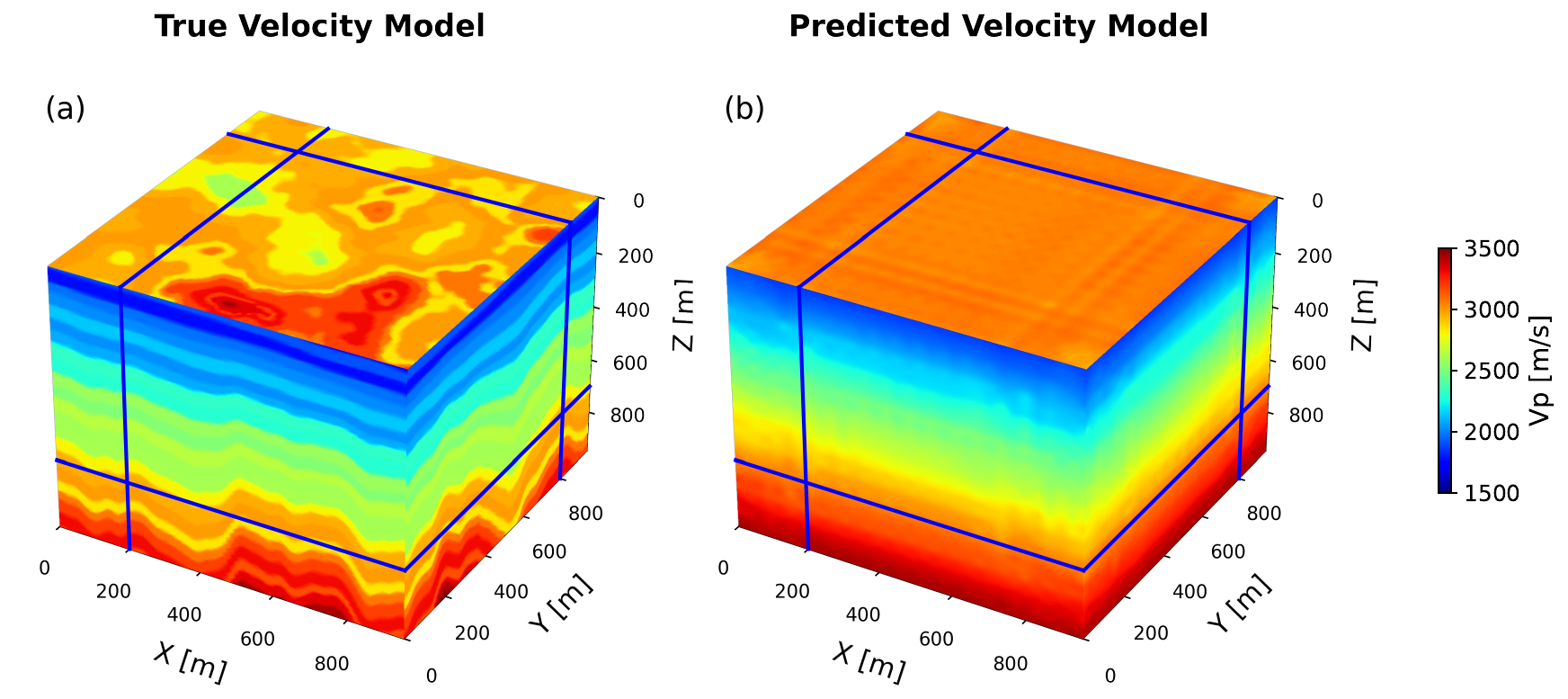}
    \caption{Preliminary evaluation of 3D end-to-end data-driven FWI. (a) The true 3D velocity model characterized by high-wavenumber stratigraphic boundaries. (b) The predicted velocity model generated directly from synthetic shot gathers by the SeisInvNet3D. While the network recovers the macroscopic, low-frequency kinematic velocity trend, it struggles to resolve sharp geological interfaces.}
    \label{fig:dlfwi_v2}
\end{figure}

\noindent
\textbf{End-to-End Data-driven Inversion:}\\
To evaluate the performance of the GeoFWI3D dataset for end-to-end mapping, we trained a 3D convolutional architecture, derived from the 2D SeisInvNet \cite{shucai2020}, to directly map volumetric velocity from raw 3D pre-stack shot gathers. Figure \ref{fig:dlfwi_v2} presents a preliminary inference result on a training volume. The network demonstrates a strong capability for macro-model building, successfully recovering the global, low-frequency kinematic velocity trend directly from the wavefield data without requiring an initial mathematical model or iterative PDE evaluations. 
However, as an end-to-end solver, the purely data-driven architecture currently exhibits distinct limitations. The predicted volume (Figure \ref{fig:dlfwi_v2}b) lacks the high-wavenumber structural resolution of the ground truth, resulting in blurred stratigraphic interfaces. 
Because the network recovers the low-frequency background trend that is difficult to obtain using standard methods, this output serves as a kinematically accurate initial model. Feeding this network prediction directly into the conventional, gradient-based acoustic FWI engine or advanced DL-assisted FWI workflow allows the physics-based optimizer to efficiently resolve the remaining high-resolution structural edges, leveraging the strengths of both data-driven initialization and PDE-based refinement.

\subsection{Data-Driven Surrogate for Traveltime and Wavefield Prediction }

Deep learning surrogates offer an alternative to iterative Partial differential equation (PDE) solvers during inference. To evaluate the applicability of the GeoFWI3D dataset for training 3D neural architectures, a 3D Fourier neural operator (FNO) \cite{li2020fourier} for kinematic traveltime prediction, and a U-Net enhanced FNO (UFNO) for spatiotemporal wavefield simulation are implemented.

Computing traveltimes through structurally complex media is a prerequisite for migration and inversion workflows. Algorithms such as the Fast marching method (FMM) \cite{sethian1999fast,pyekfmm} compute these fields by solving the Eikonal equation, which is computationally expensive when evaluated over dense 3D source grids. As an alternative, we trained a 3D FNO to approximate the traveltime solutions.  To condition the neural network training, we adopted a factorized approach, decomposing the total traveltime field ($T$) into an analytical background field ($T_0$) and a structurally dependent perturbation field ($\tau$), where $T = T_0 + \tau$. The field $T_0$ represents the straight-ray distance normalized by the velocity at the source location. The 3D FNO maps a multi-channel input comprising the normalized 3D velocity model, the analytical $T_0$ field, and a Gaussian-smoothed source-location map to the perturbation field $\tau$. 

\begin{figure}[htbp]
    \centering
    \includegraphics[width=1.0\linewidth]{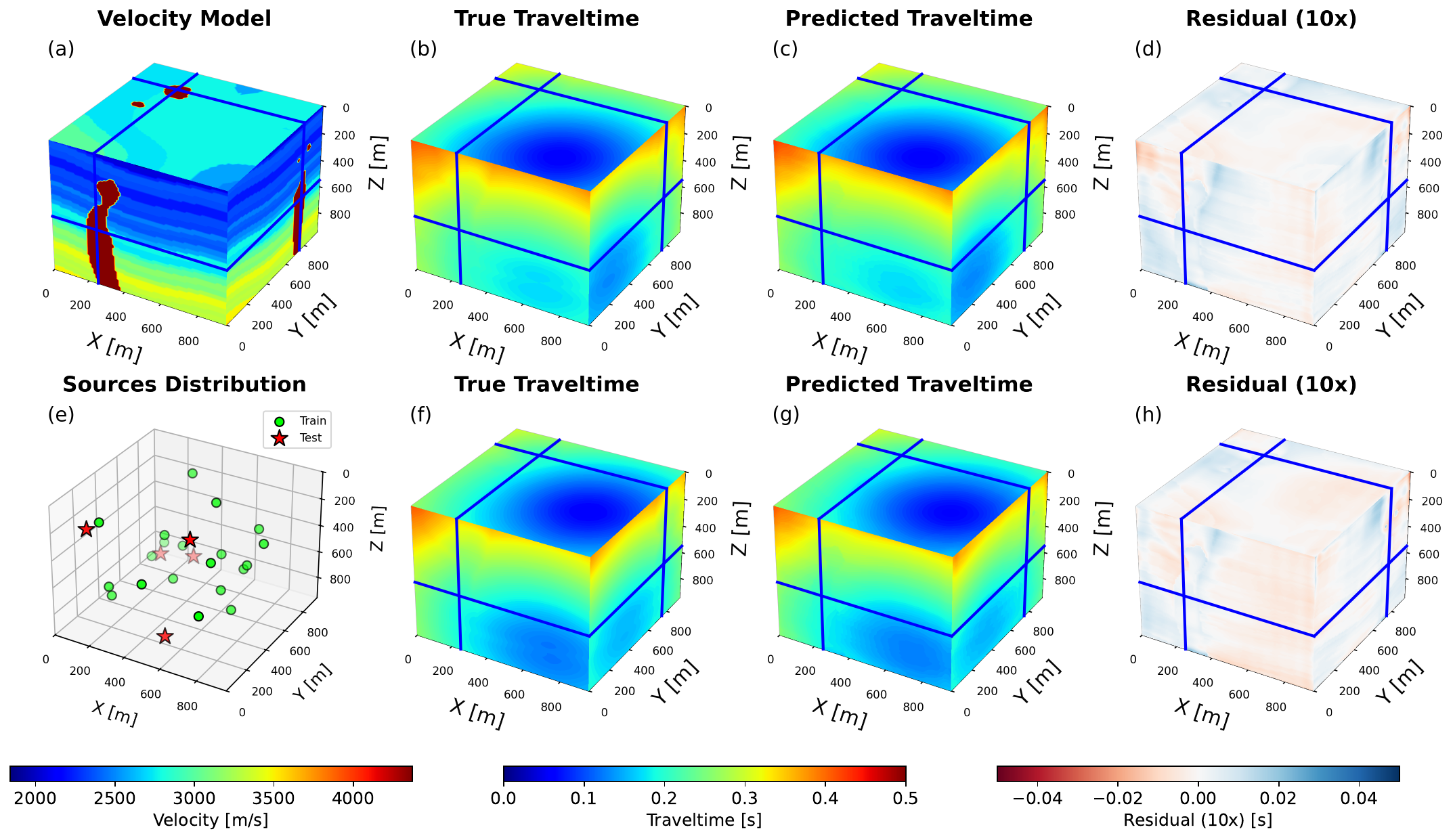}
    \caption{Performance of the 3D Fourier Neural Operator (FNO) for seismic traveltime prediction in a complex heterogeneous medium. 
    (a) 3D velocity model with salt body. 
    (b, f) Reference traveltime fields computed using the second-order fast marching method (FMM) \cite{pyekfmm} for two distinct test source locations. 
    (c, g) Corresponding traveltime volumes predicted by the trained FNO. 
    (d, h) Residual distributions ($T_{FMM} - T_{FNO}$) amplified by 10x. 
    (e) Spatial distribution of the 3D source coordinates, where green spheres denote the training set and red stars indicate the test sources. 
    Note that the highest residuals correlate with high-frequency velocity contrasts and salt boundaries, where spectral truncation in the Fourier domain limits the recovery of sharp gradient changes.}
    \label{fig:fno-tt}
\end{figure}

The FNO was trained on traveltime fields generated from 20 randomly distributed source locations and evaluated on 5 unseen sources. The network achieved an MAE of 6.836 ms and an RMSE of 9.885 ms. These values indicate that the errors are spatially stable and not dominated by large outliers, suggesting the model approximated the underlying Eikonal solution. 
Figure \ref{fig:fno-tt} illustrates the performance of the FNO on a withheld test set. The model recovers the primary kinematic trends and the wavefront distortions induced by high-contrast features, such as the fault planes and salt geometries. The absolute residual distributions (Figure \ref{fig:fno-tt}d, h) reveal that, while the global phase is well preserved, localized errors persist at sharp rheological boundaries. These residuals are likely attributable to the spectral truncation inherent in the Fourier layers, which acts as a low-pass filter on the velocity-dependent traveltime gradients. Nevertheless, the results demonstrate that the FNO architecture effectively approximates the underlying Eikonal operator for 3D heterogeneous media.

\begin{figure}[htbp]
    \centering
    \includegraphics[width=\linewidth]{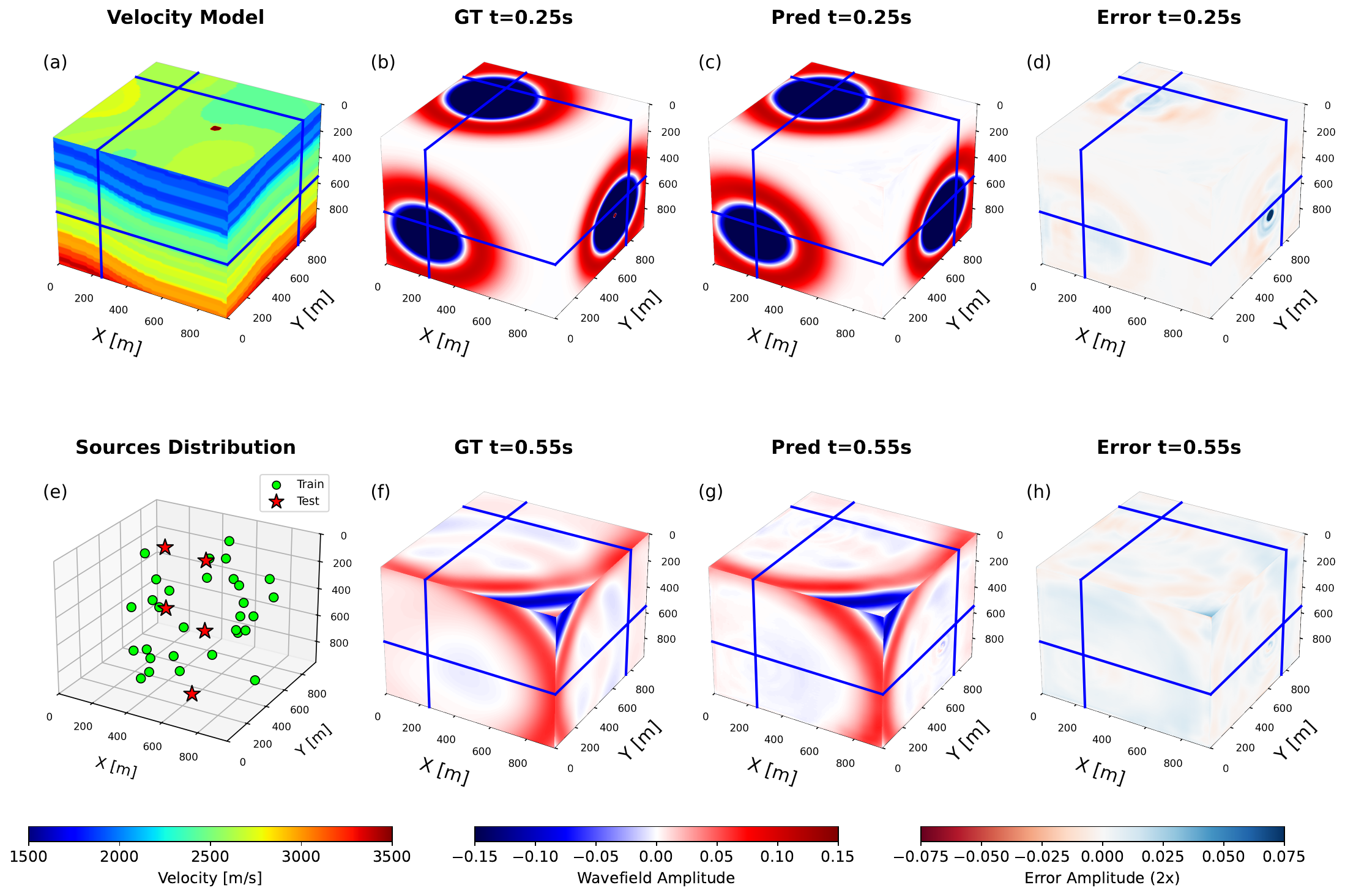} 
    \caption{3D acoustic wavefield simulation using a U-Net enhanced Fourier Neural Operator (UFNO). (a) The input 3D velocity model extracted from the GeoFWI3D dataset. (e) The spatial distribution of the randomly generated training (green circles) and testing (red stars) source locations. (b–d) The ground-truth finite-difference wavefield, the UFNO prediction, and the error at $t = 0.25$~s. (f–h) The corresponding ground-truth, prediction, and error at $t = 0.55$~s. The data-driven UFNO surrogate approximates complex wave-propagation dynamics, maintaining low residual errors even as the wavefield undergoes severe scattering due to underlying structural heterogeneities.}
    \label{fig:ufno-wavefields}
\end{figure}

To extend this validation to full dynamic wavefield propagation, we implemented a 3D acoustic wavefield surrogate using a U-Net enhanced FNO (UFNO). Because the core dataset provides structural models rather than raw wavefields, target data were generated using the Deepwave finite-difference solver on a $96 \times 96 \times 96$ velocity model with 10~m grid spacing. The acoustic wave equation was solved using a fourth-order spatial discretization with 20-cell perfectly matched layer (PML) boundaries, driven by a 5~Hz Ricker wavelet source. Each simulation spanned 1.0~s, from which 20 spatiotemporal wavefield snapshots were uniformly sampled.
The UFNO surrogate was trained to map the static velocity model and source location directly onto the complete wavefield evolution, using 30 training sources and 5 unseen test sources. As illustrated in Figure \ref{fig:ufno-wavefields}, the data-driven model accurately predicts the spatiotemporal dynamics of the propagating wavefield. Rather than approximating a smoothed kinematic trend, the network reproduces the high-frequency dynamics of the acoustic wave equation, including phenomena such as diffraction and multipathing. The errors remain low throughout the simulation, suggesting that the surrogate preserves both phase coherence and amplitude fidelity after the wavefront is scattered by the underlying structural heterogeneities.

\subsection{Training 3D Diffusion Models}
\begin{figure}
    \centering
    \includegraphics[width=0.99\linewidth]{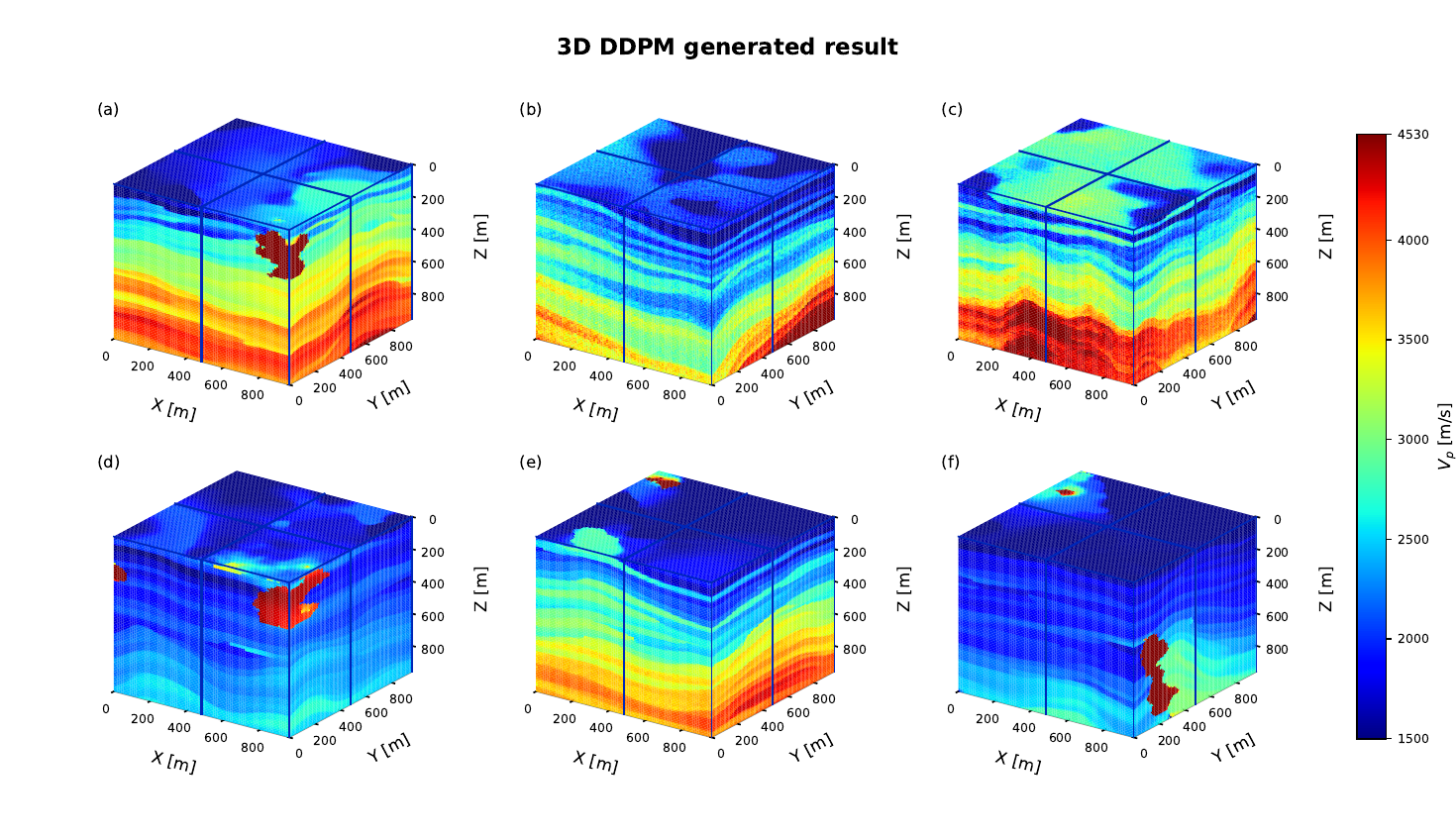}
    \caption{Unconditional 3D velocity models ($V_p$) generated by the baseline Denoising Diffusion Probabilistic Model (DDPM) trained on a subset of the GeoFWI3D dataset. Panels (a) through (f) display independent, randomly generated realizations. The diffusion model captures the primary geological distributions, synthesizing plausible background stratigraphic folding and high-velocity salt intrusions (dark red) while adhering to the native $96 \times 96 \times 96$ spatial dimensions and physical velocity bounds.}
    \label{fig:diffusion3d}
\end{figure}

As a proof-of-concept generative benchmark, we trained a compact three-dimensional denoising diffusion probabilistic model (DDPM) \cite{ho2020denoising} directly on the raw GeoFWI3D $V_p$ volumes. To keep the experiment computationally feasible for rapid demonstration, 2,000 models were randomly selected from the 10,000-volume repository and split into 1,800 training and 200 validation samples. Each velocity volume was kept at the native $96 \times 96 \times 96$ resolution and normalized to $[-1,1]$ using the global dataset velocity range of approximately 1,500--4,530~m/s. The denoising network was a 3D U-Net with channel widths $(32,64,128,256)$, attention applied at the two coarsest scales, and two residual blocks per level. The model was optimized with AdamW using a learning rate of $10^{-4}$, weight decay of $10^{-4}$, a batch size of 1, and gradient accumulation over 4 steps. An exponential moving average (EMA) of the model weights with decay 0.999 was maintained for sampling. The diffusion process used 1,000 training noise steps with a scaled linear beta schedule, while qualitative samples were generated from the EMA weights using DDIM sampling with 250 inference steps. 
Figure \ref{fig:diffusion3d} shows that the DDPM synthesizes geologically plausible 3D velocity volumes. The generated models reproduce the primary macroscopic features of the training distribution, including conformal sedimentary folding (e.g., Figures \ref{fig:diffusion3d}b, e) and high-contrast, structurally complex salt diapirs (e.g., Figures \ref{fig:diffusion3d}a, f). While the generated sharp interfaces exhibit minor volumetric smoothing, typical of diffusion-based synthesis, the preservation of both continuous velocity gradients and discrete structural anomalies demonstrates that the dataset provides sufficient statistical variance and spatial coherence to support robust 3D generative deep learning benchmarks.

\subsection{Limitations of GeoFWI3D Dataset}

While GeoFWI3D provides a large-scale, structurally diverse benchmark for 3D seismic ML, it has several inherent limitations characteristic of synthetic datasets and should be carefully considered by practitioners before applying models trained on it to real-world seismic workflows.  The petrophysical models and their corresponding seismic responses are grounded in the acoustic approximation. Real-world subsurface media are fundamentally elastic/viscoelastic, highly anisotropic. Hence, the dataset does not capture complex elastic properties or intrinsic attenuation, which are often critical for advanced multi-parameter inversion. Such advanced tasks should not solely rely on the GeoFWI3D dataset to transfer directly to complex field data without domain adaptation. A fully elastic multiparameter model that accounts for anisotropy and attenuation will be the focus of our future study.

The synthetic seismic volumes provided in the dataset were generated by convolving the broadband reflectivity with a 3D Point Spread Function in the wavenumber domain, rather than by a full finite-difference wavefield simulation. For FWI workflows, users ought to generate their own physically accurate synthetic shot gathers by deploying finite-difference or finite-element wave-equation solvers on the provided raw velocity volumes. Geological features that naturally occur at lower resolution scale such as sub-seismic fracture swarms, localized diagenetic alterations, and thinly laminated thin-bed sequences—are not explicitly resolved in the velocity models.  While the RGM framework produces geometrically diverse and visually plausible subsurface models, the generation process is fundamentally geometric rather than process-based. True geological structures emerge from coupled physical processes such as differential compaction, tectonic stress evolution, salt dissolution and reprecipitation, and fault growth by segment linkage. Individual model realizations are therefore geologically plausible in aggregate but are not conditioned on the depositional histories or stress regimes of specific regional geology.

These limitations define a clear scope for the appropriate use of GeoFWI3D, which is best suited as a controlled benchmark dataset for evaluating the structural sensitivity, generalization capacity, and computational scalability of 3D seismic ML architectures under acoustic conditions. It is not intended as a substitute for field-calibrated training data in production workflows, and results obtained on GeoFWI3D should be interpreted as indicative rather than predictive of real-world algorithm performance.

\section{Usage Notes}
The full dataset is hosted at \url{https://doi.org/10.5281/zenodo.20148778} \cite{swaminadhan_2026_20148778} and distributed under Creative Commons Attribution 4.0 International.
Navigate to the link provided in the repository and download the required \texttt{models\_batch\_*.tar.gz} archive files to your local machine or compute cluster.
Once downloaded, place all the \texttt{.tar.gz} files in a single directory. You can extract all batches into a unified \texttt{allmodels} directory using the following bash script:
\begin{lstlisting}[language=bash, caption=Bash script to extract all model batches into a single directory.]
#!/bin/bash

mkdir -p allmodels

for archive in models_batch_*.tar.gz; do
    echo "Extracting $archive into allmodels/..."
    tar -xzf "$archive" -C allmodels
done
\end{lstlisting}

The dataset volumes are stored as raw 32-bit float binary files (\texttt{.bin}). We strongly recommend using \texttt{numpy.memmap} to read the files. Below is a ready-to-use Python function to load the seismic amplitudes, velocity model, relative geologic time (RGT), fault model, and derive the salt mask for a given model index. The returned arrays follow an \texttt{(X, Y, Z)} index order.

\begin{lstlisting}[language=Python, caption=Python script to load 3D binary volumes into memory-mapped arrays.]
import os
import numpy as np

def read_models(data_root, model_folders, index):
    """
    Loads 3D volumes from binary files for a specific model index.
    The returned arrays have an index order of (X, Y, Z).
    """
    # Define file paths
    # Seismic reflectivity
    image_file = os.path.join( 
        data_root, 
        model_folders[index], 
        "image3d.bin"
    ) 
    # Velocity model
    vp_file = os.path.join(
        data_root, 
        model_folders[index], 
        "vp3d.bin"
    )
    # Relative geologic time
    rgt_file = os.path.join(
        data_root, 
        model_folders[index], 
        "rgt3d.bin"
    )
    # Fault model
    fault_file = os.path.join(
        data_root, 
        model_folders[index], 
        "fault3d.bin"
    ) 

    # The dimensions are ordered (X, Y, Z)
    nx, ny, nz = 96, 96, 96
    shape = (nx, ny, nz)

    # Load volumes using memory mapping
    image = np.memmap(
        image_file, 
        dtype=np.float32, 
        mode='r', 
        shape=shape
    )
    vp = np.memmap(
        vp_file, 
        dtype=np.float32, 
        mode='r', 
        shape=shape
    )
    rgt = np.memmap(
        rgt_file, 
        dtype=np.float32, 
        mode='r', 
        shape=shape
    )
    fault = np.memmap(
        fault_file, 
        dtype=np.float32, 
        mode='r', 
        shape=shape
    )

    # Derive salt mask: Salt bodies are represented by an RGT value of 0
    salt = (rgt == 0)
    salt_mask = salt.astype(np.float32)

    return image, vp, rgt, fault, salt_mask
\end{lstlisting}

To facilitate training deep learning architectures, the memory-mapped arrays must be encapsulated into framework-specific data pipelines. The following listing provides a PyTorch \texttt{Dataset} and \texttt{DataLoader} implementation designed specifically for the fault detection example. It directly reads raw binary files, maintains the native \texttt{(X, Y, Z)} spatial index order, and expands the arrays to include the single-channel dimension \texttt{(C, X, Y, Z)} required by 3D convolutional operators.

\begin{lstlisting}[language=Python, caption=PyTorch Dataset and DataLoader implementation for 3D fault segmentation.]
import os
import numpy as np
import torch
from torch.utils.data import Dataset, DataLoader

class GeoFWI3DFaultDataset(Dataset):
    """
    PyTorch Dataset for 3D fault detection.
    Loads raw 32-bit float binary volumes directly into tensors.
    """
    def __init__(self, data_root, model_folders):
        self.data_root = data_root
        self.model_folders = model_folders
        # The native spatial dimensions are ordered (X, Y, Z)
        self.shape = (96, 96, 96)

    def __len__(self):
        return len(self.model_folders)

    def __getitem__(self, idx):
        # Define specific file paths for the input and target
        image_file = os.path.join(
            self.data_root, 
            self.model_folders[idx], 
            "image3d.bin"
        )
        fault_file = os.path.join(
            self.data_root, 
            self.model_folders[idx], 
            "fault3d.bin"
        )

        # Load volumes using memory mapping
        image_memmap = np.memmap(
            image_file, 
            dtype=np.float32,
            mode='r', 
            shape=self.shape
        )
        fault_memmap = np.memmap(
            fault_file, 
            dtype=np.float32, 
            mode='r', 
            shape=self.shape
        )

        # Copy into memory to avoid PyTorch writable 
        # warnings with memmap arrays
        image_array = np.array(image_memmap)
        fault_array = np.array(fault_memmap)

        # convert fault indices to fault mask
        fault_mask = (fault_array > 0).astype(np.float32)
        

        # Convert to PyTorch tensors and expand to include 
        # the channel dimension (1, X, Y, Z)
        image_tensor = torch.from_numpy(image_array).unsqueeze(0)
        fault_tensor = torch.from_numpy(fault_mask).unsqueeze(0)

        return image_tensor, fault_tensor

# Example usage for initializing the DataLoader:
dataset = GeoFWI3DFaultDataset(
    data_root="./allmodels/", 
    model_folders=[f"model_{i:04d}" for i in range(16)]
)
train_loader = DataLoader(
    dataset, 
    batch_size=4, 
    shuffle=True, 
    num_workers=4
)
\end{lstlisting}

\section{Code availability}
The quick start guides and baseline benchmarking scripts are made public and can be accessed at \url{https://github.com/aaspip/geofwi3d}.

\section*{Acknowledgments}
The authors would like to thank the Texas Consortium of Computational Seismology (TCCS) for sponsoring the research. The authors also thank Texas Seismological Network and Seismology Research (TexNet), and the State of Texas, which provided financial support for this publication under the University of Texas at Austin award \# 201503664. K.G. and T.C. are supported by Los Alamos National Laboratory (LANL) Laboratory Directed Research and Development (LDRD) project \# 20240322ER.

\section*{Author contributions}
S.S., Y.S., C.L., S.F., and Y.K.C. conceptualized the research idea. S.S., K.G., and T.C. contributed to the creation of the initial GeoFWI3D dataset. S.S., Y.S., and C.L. conducted most of the data preparation, processing, benchmarking, and analysis. J.L., and R.D. contributed to the realism of the velocity model.  T.A., Y.C., L.Y., and U.W. assisted with validation, experiments, and analysis of results. S.S., Y.S., C.L., L.Y., and M.K.S. contributed to the FWI experiments. Y.C. and U.W. contributed to the scientific ML application of GeoFWI3D.  S.S., C.L., and Y.K.C. participated in visualization. S.S and Y.K.C. wrote the initial manuscript.  All authors reviewed and approved the final manuscript.

\section*{Competing interests}
The authors declare no competing interests.

\bibliographystyle{unsrt}
\bibliography{nat,tomo}

\end{document}